\documentclass[a4paper,superscriptaddress,nofootinbib]{revtex4-1}

\usepackage{epsfig}
\usepackage{amsfonts}
\usepackage{amsmath}
\usepackage{amssymb}
\usepackage{amsthm}
\usepackage{array}
\usepackage{booktabs}
\usepackage{color}
\usepackage{graphicx}
\usepackage{hyperref}
\usepackage{multirow}
\usepackage{slashed}
\usepackage{subfigure}
\usepackage{theorem}
\usepackage{textcomp}

\allowdisplaybreaks[1]

\numberwithin{equation}{section}

\newcommand{\be}{\begin{equation}}
\newcommand{\ee}{\end{equation}}
\newcommand{\beq}{\begin{equation}}
\newcommand{\eeq}{\end{equation}}
\newcommand{\bea}{\begin{eqnarray}}
\newcommand{\eea}{\end{qnarray}}

\def\<{\left\langle}
\def\>{\right\rangle}

\begin{document}

\title{Photon-initiated production of a di-lepton final state at the LHC:\\
cross section versus forward-backward asymmetry studies}

\author{Elena Accomando}
\email[E-mail: ]{E.Accomando@soton.ac.uk}
\affiliation{School of Physics \& Astronomy, University of Southampton,
        Highfield, Southampton SO17 1BJ, UK}
\affiliation{Particle Physics Department, Rutherford Appleton Laboratory, 
       Chilton, Didcot, Oxon OX11 0QX, UK}

\author{Juri Fiaschi}
\email[E-mail: ]{Juri.Fiaschi@soton.ac.uk}
\affiliation{School of Physics \& Astronomy, University of Southampton,
        Highfield, Southampton SO17 1BJ, UK}
\affiliation{Particle Physics Department, Rutherford Appleton Laboratory, 
       Chilton, Didcot, Oxon OX11 0QX, UK}
       
\author{Francesco Hautmann}
\email[E-mail: ]{hautmann@thphys.ox.ac.uk}
\affiliation{Particle Physics Department, Rutherford Appleton Laboratory, 
       Chilton, Didcot, Oxon OX11 0QX, UK}
\affiliation{Theoretical Physics Department, University of Oxford, Oxford OX1 3NP}

\author{Stefano Moretti}
\email[E-mail: ]{S.Moretti@soton.ac.uk}
\affiliation{School of Physics \& Astronomy, University of Southampton,
        Highfield, Southampton SO17 1BJ, UK}
\affiliation{Particle Physics Department, Rutherford Appleton Laboratory, 
       Chilton, Didcot, Oxon OX11 0QX, UK}

\author{Claire H. Shepherd-Themistocleous}
\email[E-mail: ]{claire.shepherd@stfc.ac.uk}
\affiliation{School of Physics \& Astronomy, University of Southampton,
        Highfield, Southampton SO17 1BJ, UK}
\affiliation{Particle Physics Department, Rutherford Appleton Laboratory, 
       Chilton, Didcot, Oxon OX11 0QX, UK}

\begin{abstract}
\noindent
We explore the effects of Photon Induced (PI) production of a dilepton final state in the Large Hadron Collider environment. Using QED Parton Distribution Function (PDF) sets we can treat the photons as real partons inside the protons and compare their yield directly to that of the Drell-Yan (DY) process. In particular, we concentrate on an error analysis of the two mechanisms. 
In order to do so, we use the NNPDF set, which comes with a set of replicas to estimate the systematic PDF error, and the CT14 set. On the one hand, we find that the PI contribution becomes dominant over DY above a dilepton invariant mass of 3 TeV. On the other hand, the PI predictions are affected by a large uncertainty coming from the QED PDFs, well above the one affecting the DY mode. We assess the impact of these uncertainties in the context of resonant and non-resonant searches for a neutral massive vector boson ($Z^\prime$) through the differential cross section and Forward-Backward Asymmetry (AFB) observables as a function of the dilepton invariant mass. While the former is subject to the aforementioned significant residual errors the latter shows the systematic error cancellation expected (recall that AFB is a ratio of cross sections) even in presence of PI contributions, so that the recently emphasized key role played by AFB as a valid tool for both $Z^\prime$ discovery and interpretation in both resonant and non-resonant mode is further consolidated.
\end{abstract}

\pacs{NN.NN.NN.Cc}

\maketitle

\tableofcontents

\setcounter{footnote}{0}

\section{Introduction}\label{sec:intro}

The upgrade in energy of the Large Hadron Collider (LHC) to 13 TeV has opened the exploration of higher energy scales that were barred during the past RunI. A crucial point in the search for Beyond Standard Model (BSM) physics is the precise understanding of the behaviour of the SM background, especially in the high energy regime where one expects that new physics could appear. The LHC potential in BSM searches at the ongoing RunII will be further boosted by the increase of the collected data sample in three years time when the luminosity should reach the project value $L=300~fb^{-1}$. As we approach the high luminosity phase at the LHC, the statistical errors will be greatly reduced. At the same time, systematic effects will become more and more important. In a hadron collider like the LHC, the major source of theoretical systematics at high energies comes from the PDF uncertainties. The errors in measured data propagate in fact into the fitted PDFs. Great improvements in this sense have been achieved recently by many PDF collaborations. The parameterization and modeling of quarks and gluons PDFs have been significantly ameliorated by including also new high precision data from HERA and Fermilab (see Ref. \cite{Accardi:2016qay, Accardi:2016ndt} and references therein). This has led to a reduction of the uncertainties on the $d$/$u$ ratio, especially in the large-$x$ regime ($x\ge 0.4$) closely related to the high energy scales probed in parton-parton hard scatterings. 
 
\noindent
The advantage of such an improvement easily translates into an enhanced capability of producing more accurate predictions for BSM signals and SM background at the LHC. New physics signals require an accurate determination of the SM background and uncertainties in the large-$x$ PDFs could affect the interpretation of the LHC experiments searching for new particles at high mass scales. In particular, we consider in this paper the leptonic Drell-Yan (DY) channel $pp\rightarrow q\bar q\rightarrow l^+l^-$ with $l=e, \mu$. This process is particularly useful in the search for extra heavy neutral spin-1 $Z^\prime$-bosons. For a review see Refs. \cite{Langacker:2008yv,Erler:2009jh,Nath:2010zj,Accomando:2010fz,Accomando:2013sfa,Accomando:2016sge,Accomando:2016mvz,Belyaev:2008yj} and references therein.  As the present bounds on the $Z^\prime$-boson mass, extracted at the past LHC RunI, have been set at around $M_{Z'}\simeq 2500 - 3200$ GeV depending on the specific theoretical model \cite{Accomando:2015cfa, Khachatryan:2014fba}, the dilepton spectrum of interest at the ongoing LHC RunII lies at high mass scales. To a large extent, its simulation thus requires to have large values of the fraction of longitudinal proton momentum taken by the colliding quarks and antiquarks initiating the hard scattering for the $Z^\prime$-boson production in the DY channel. For this reason, the improved knowledge of the quark and antiquark PDFs at large-$x$ is extremely valuable. 

\noindent
With increasing the luminosity towards the LHC project value $L=300~fb^{-1}$, the statistical error will get smaller and smaller at medium-large energy scales while higher energy scales will be explored for the first time saturating the LHC potential in discovering (or excluding) new physics. It is mandatory that theoretical uncertainties follow this trend too, for the interpretation of the experimental results that will be obtained with a very good statistical precision in many cases. Fixed-order perturbative QCD calculations for Drell-Yan production are available at next-to-leading (NLO) and next-to-next-to-leading (NNLO) accuracy, as well as EW NLO corrections for the complete di-lepton channel~\cite{Dittmaier:2009cr}.
In order to be consistent with the partonic matrix elements, the PDF sets should have both QCD and EW corrections in the Dokshitzer-Gribov-Lipatov-Altarelli-Parisi (DGLAP) evolution kernels \cite{deFlorian:2016gvk,deFlorian:2015ujt}. To stay with QED effects, in addition to the corrected kernels, the QED collinear singularity leads to the need of introducing the photon distribution function which mixes with the quark (antiquark) PDFs and requires to be determined by a fit to the experimental data, like the other PDFs. 

\noindent
This latter element can lead to a novel source of theoretical systematics. One should in fact consider the Photon Induced (PI) lepton pair production, $pp\rightarrow \gamma\gamma + X\rightarrow l^+l^- +X$ with $l=e, \mu$. This contribution sums to the DY (differential) cross section and modifies the prediction of the SM background that we have up to now. The PI dilepton production mode receives contributions from three different processes, distinguished by the virtuality of the two initial photons giving rise to the $\gamma\gamma$ hard scattering. 
When the two photons are both considered as proton constituents, one has the so-called double dissociative process. In this instance, the photon's virtuality is null so that the photon can be seen as real (resolved photon). When one photon is resolved, thus being described by a QED PDF, and the other one is emitted from a quark (antiquark) with a non-zero virtuality one has the so-called single dissociative process. The last contribution represents the case when both photons are radiated off quarks (or antiquarks) and have non-zero virtuality. The Equivalent Photon Approximation (EPA), described in Ref. \cite{Budnev:1974de}, provides a method to treat the case of non-zero virtuality.  Double and single dissociative processes require the knowledge of the QED PDFs.

\noindent
The effect of the photon induced contribution has been evaluated for different processes.  Recently, it has been analysed in the context of Higgs boson measurements. In particular, it has been computed for the four-lepton final state as a background to the Higgs production where the Higgs boson decays into a $Z$-boson pair giving rise ultimately to four leptons in the final state \cite{Dyndal:2015hrp}. The PI contribution has been also calculated for two-leptons and two-bosons final states \cite{Ababekri:2016kkj,Csaki:2016raa,Fichet:2015vvy,Harland-Lang:2016lhw,Harland-Lang:2016apc,Harland-Lang:2016qjy,Luszczak:2015aoa,daSilveira:2014jla,Luszczak:2014mta}. High-energy QCD effects in the coupling of PI processes to jets, for low but finite photon virtualities, have been studied in Refs.  \cite{Hautmann:1996zb,Bartels:1996ke,Brodsky:1996sg,Brodsky:1997sd,Hautmann:1997hh,Boonekamp:1998ve}.

\noindent
Some of the PDF collaborations have released sets that include the photon as an additional parton inside the hadron: MRST2004QED \cite{Martin:2004dh}, NNPDF2.3QED \cite{Ball:2013hta} and CT14QED \cite{Schmidt:2015zda}. Quark (antiquark) PDFs are usually accompanied by an estimate of the PDF error.  Some collaborations have implemented the Hessian procedure \cite{Stump:2003yu}, others the replicas method \cite{Ball:2011gg}. Concerning the PDF sets with QED effects, only NNPDF2.3 provides uncertainty estimates. 

\noindent
In this paper, we focus on the double dissociative process and use the NNPDF2.3QED release, which adopts the replicas prescription to calculate the PDF error, to compute the PI contribution to the dilepton final state when in presence of an hypothetical new physics signal like an extra $Z^\prime$-boson. In order to quantify the impact of the PI contribution with respect to the DY process, taking into account its theoretical uncertainty, we analyze two physical observables: the differential cross section and the forward-backward asymmetry (AFB) distribution in the dilepton invariant mass. This choice is motivated by the observation that, for $Z^\prime$-boson searches in the DY channel, AFB is much more robust against systematics than the cross section in the dilepton mass spectrum and can be used in both the discovery and interpretation stages of the experimental analysis \cite{Accomando:2015cfa,Accomando:2015pqa,Accomando:2015sun,Accomando:2015ava,Dittmar:1996my}.
A measurement of the AFB of Drell-Yan lepton pairs as a function of dilepton invariant mass and rapidity at the 7 and 8 TeV LHC has been performed by the LHC collaborations. Latest results are reported in Ref. \cite{Khachatryan:2016yte}.

\noindent
The paper is organized as follows. In Sect.~\ref{sec:PhotonPDF}, we introduce the photon induced process we are considering and we compute its contribution to the dilepton final state at both the LHC RunI and RunII, showing that the upgrade in energy (and luminosity) is a sufficient motivation for the inclusion of this process in the experimental analyses. In Sect.~\ref{sec:PDFerror}, we give a brief description of the replicas method employed by NNPDF2.3QED and we show our result for the PDF uncertainty on the dilepton mass spectrum mapped in both cross section and AFB. In Sect.~\ref{sec:cuts}, with a view to suppressing the SM background to $Z^\prime$-boson searches, we study the effect of various kinematical cuts on the PI and  DY contributions to the dilepton final state. In Sect.~\ref{sec:BSM}, we present two possible benchmark models predicting one extra heavy $Z^\prime$-boson: $E_6^\chi$ and Sequential SM (SSM). The first one belongs to the class of the $E_6$ models \cite{Langacker:2008yv,Erler:2009jh,Nath:2010zj,Accomando:2010fz} and is characterized by a narrow width $Z^\prime$-boson. The latter, taken often as benchmark by the experimental collaborations, provides an example of wide $Z^\prime$-boson \cite{Altarelli:1989ff}. 
Here we focus on the magnitude of the theoretical error coming from the PDF uncertainty on both cross section and AFB. In Sect.~\ref{sec:summary}, we conclude.

\section{Real photons from QED PDFs}\label{sec:PhotonPDF}

\begin{figure}[t]
\centering
\includegraphics[width=0.70\textwidth]{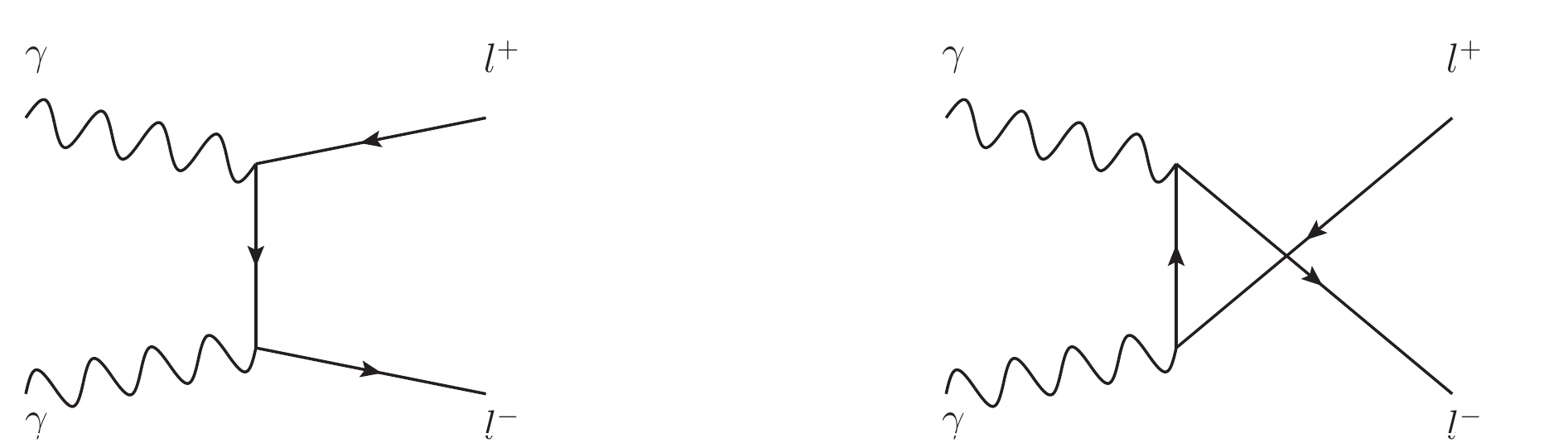}
\caption{Photon Induced process contributing to the dilepton final state.}
\label{fig:PI_digrams}
\end{figure}

In this section, we discuss the contribution to dilepton final states coming from two initial resolved photons inside the protons. The hard scattering process induced by a photon-photon collision and leading to the production of two opposite-sign leptons in the final state is mediated by the exchange of a charged lepton in the $t$- and $u$- channels. 
The lowest order Feynman diagrams contributing to this process are shown in Fig. \ref{fig:PI_digrams}. 
The kinematics of the photon induced scattering has been extensively studied in the literature. 
Its lowest order matrix element squared at parton level can easily be computed as \cite{Halzen:1984mc}
\begin{equation}
\left|\mathcal{M}(\gamma\gamma\rightarrow l^+l^-)\right|^2 = 2e^4 \left(\frac{t}{u}+\frac{u}{t}\right), 
\end{equation}
where $u$ and $t$ are the usual Mandelstam variables and $e$ is the electron electric charge.

\noindent
Recently, three PDF collaborations, MRST2004QED, NNPDF2.3QED and CT14QED, have released their PDF sets that include the photon as a proton constituent. This means that those photons extracted from the proton are on-shell ($q_\gamma^2 = 0$), i.e., they are real photons. In the literature, it has also been studied the contribution to dilepton final states coming from one or two quasi-real photons radiated off quarks (antiquarks) inside the proton. This contribution has been calculated adopting the EPA. Following this prescription, one can evaluate the effect of quasi-real photons ($q_\gamma^2 \approx 0$), neglecting the contribution of very off-shell photons which is anyhow subdominant. This prescription has been implemented in some of the most popular computational tools for particle physics: Pythia8 \cite{Sjostrand:2007gs} and Madgraph 5.1 \cite{Alwall:2011uj}. 

\noindent
We address in this paper the PI contribution to the dilepton final state where the real photons come from the QED PDFs, postponing the discussion of our results on the other two contributions computed in EPA to a forthcoming publication. We compute the double dissociative process using the aforementioned three different PDF sets. In all cases, we fix the factorization scale to be equal to the dilepton invariant mass, $Q = M_{ll}=\sqrt{\hat s}$. We have verified that a different scale choice, for example $Q^2 = p_T^2$, where $p_T$ is the lepton transverse momentum, does not significantly change our conclusions. The integration of the PI matrix element squared is collinearly divergent. However, the implementation of the usual acceptance cuts can regularize the computation. We thus impose that $|\eta_l| < 2.5$ and $p_T^l > 20~GeV$ where $\eta_l$ and $p_T^l$ are the rapidity and transverse momentum of each final state lepton, respectively.
In the following we will use the NNPDF2.3QED results unless otherwise specified.

\begin{figure}[t]
\centering
\includegraphics[width=0.45\linewidth]{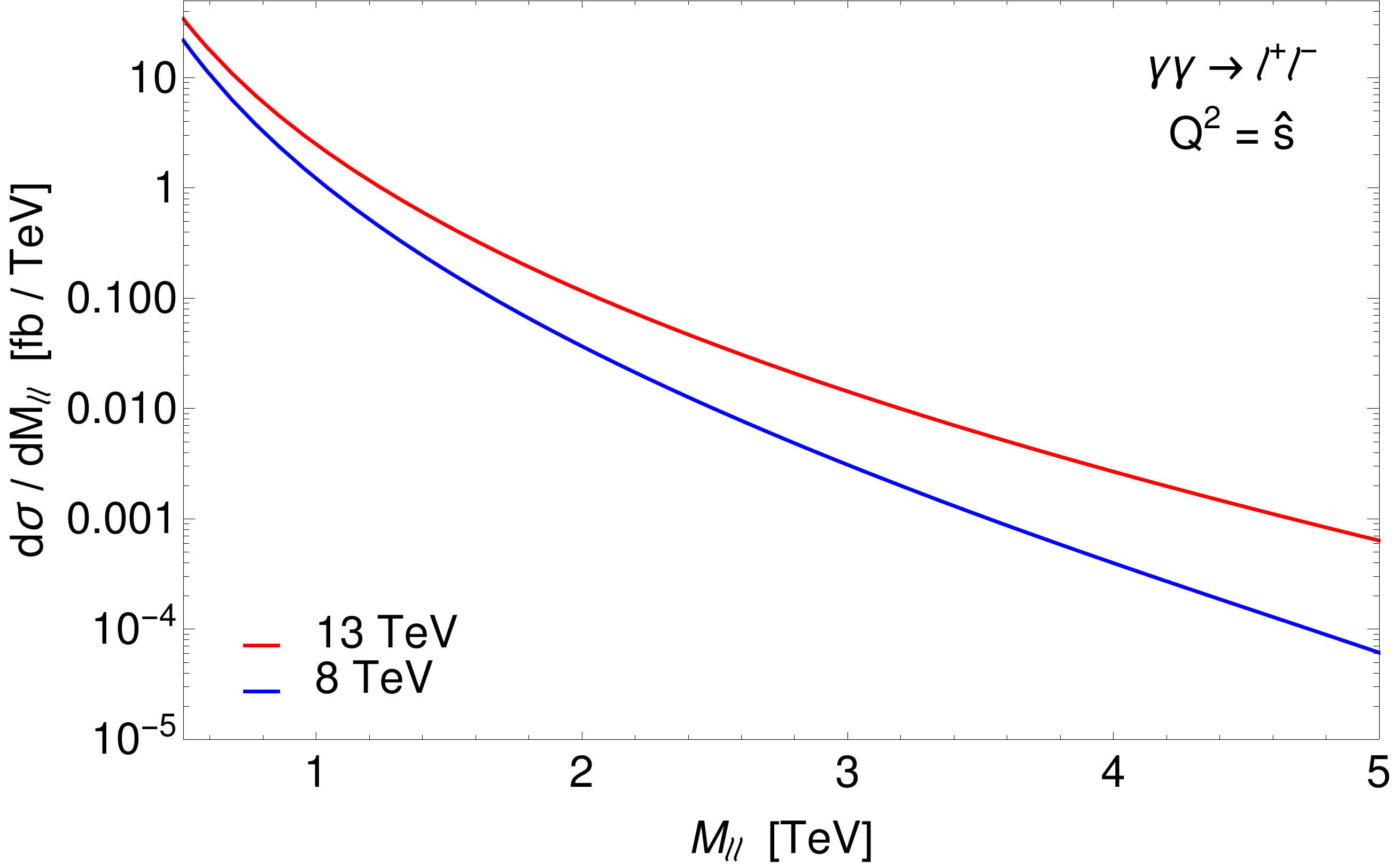}{(a)}
\includegraphics[width=0.45\linewidth]{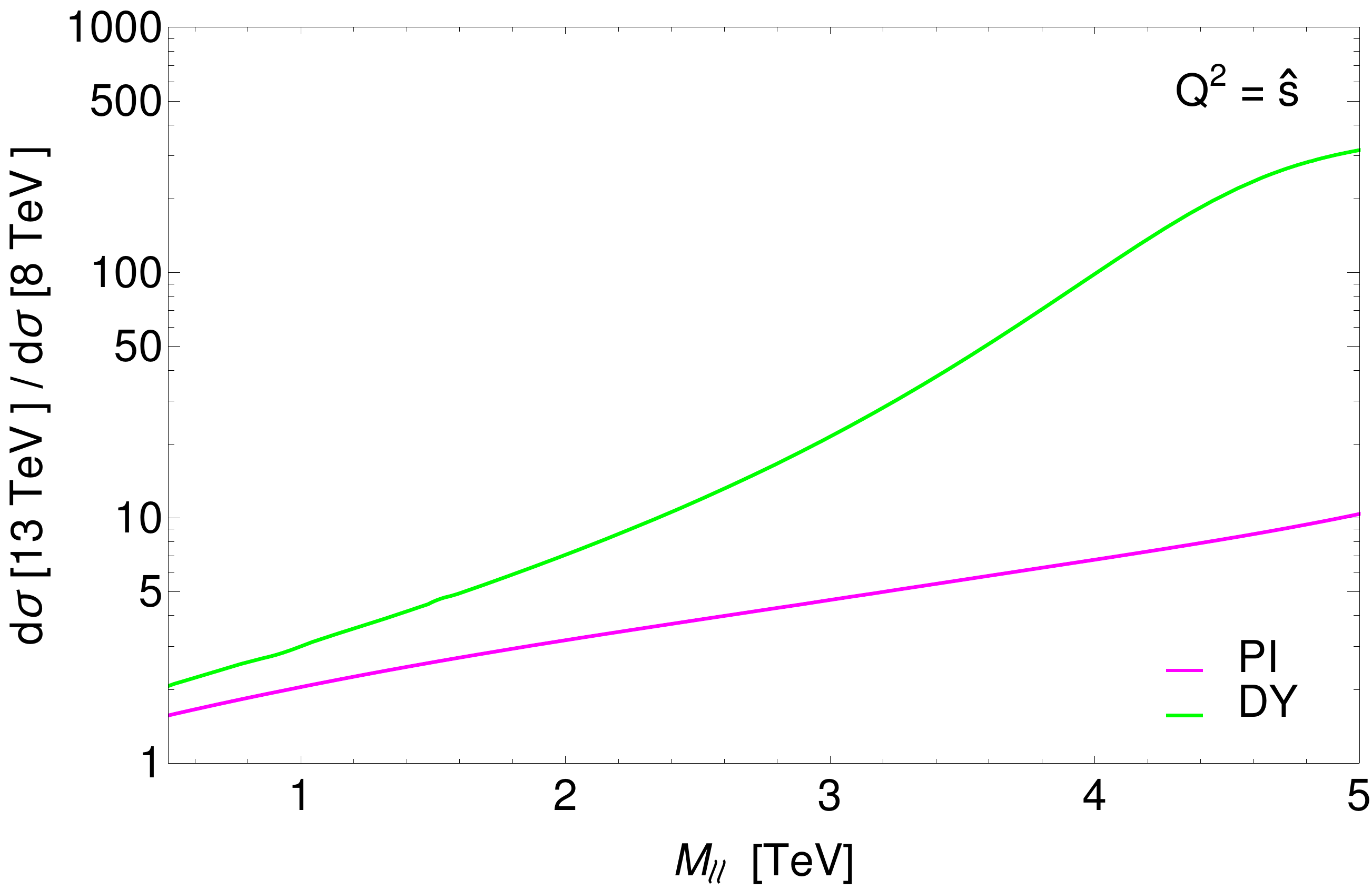}{(b)}
\caption{(a) PI differential cross section in the dilepton invariant mass for the LHC at 8 TeV (blue line) and 13 TeV (red line). 
(b) Ratio of the two differential cross sections at 13 TeV and 8 TeV (magenta line) compared with the analogous ratio for the DY case (green line).
Standard acceptance cuts are applied: $|\eta_l| < 2.5$ and $p_T^l > 20~$GeV.}
\protect{\label{fig:13vs8}}
\end{figure}

\begin{figure}[t]
\centering
\includegraphics[width=0.45\linewidth]{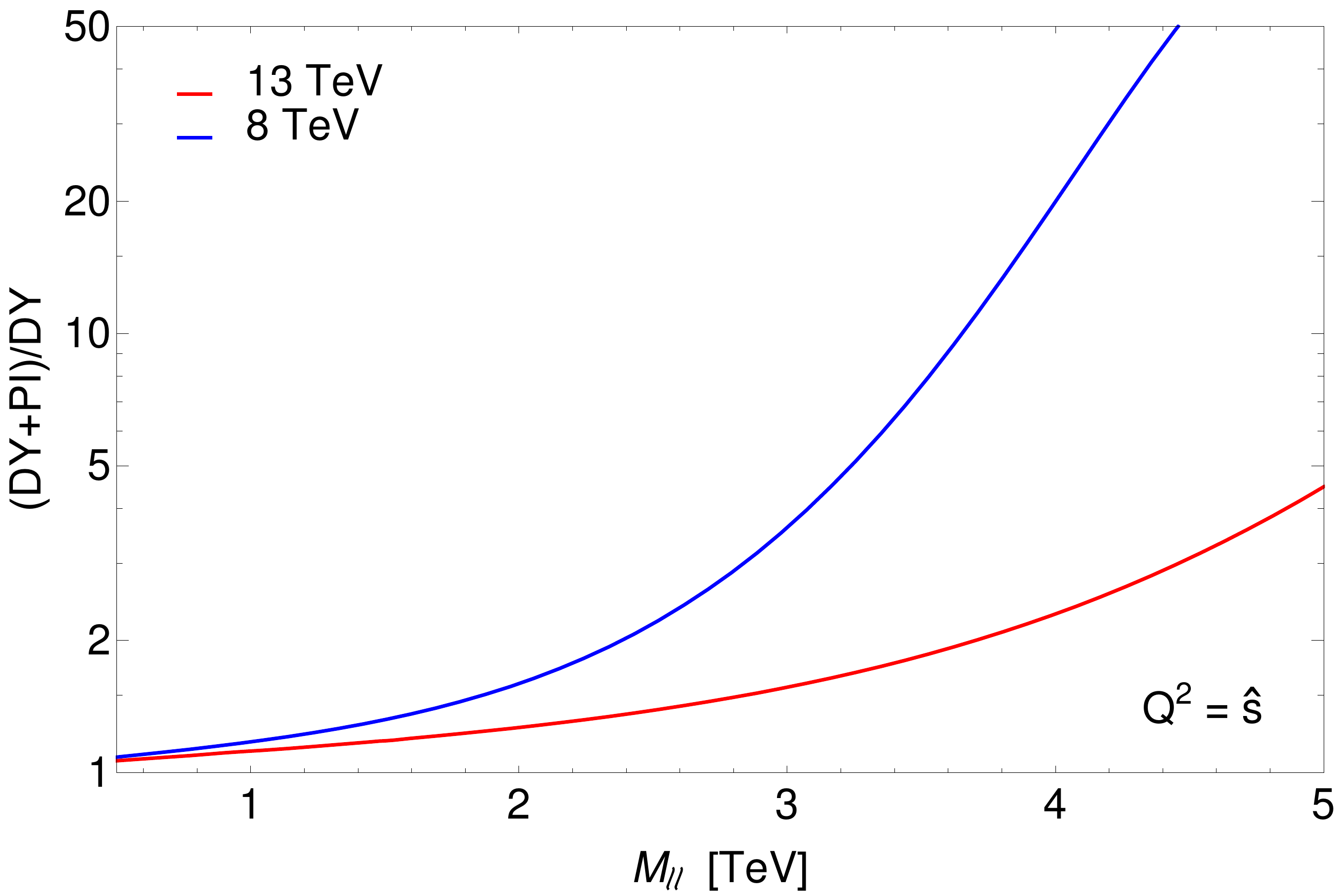}{(a)}
\includegraphics[width=0.45\linewidth]{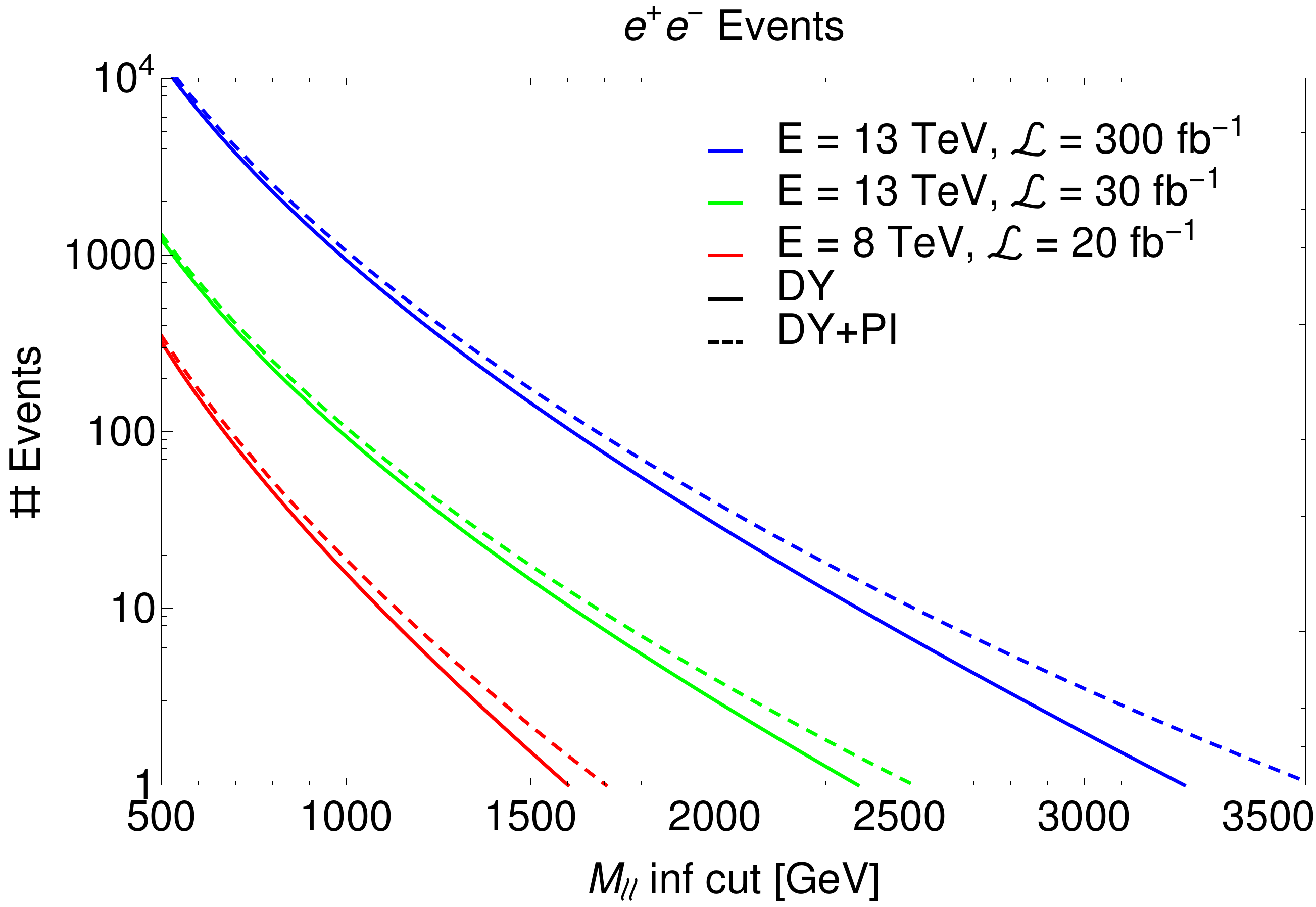}{(b)}
\caption{(a) Relative importance of the PI term in the dilepton SM background at the LHC with energies of 8 TeV (blue line) and 13 TeV (red line). Standard acceptance cuts are applied ($|\eta_l| < 2.5$ and $p_T^l > 20~GeV$).
(b) Number of events expected in the dielectron channel as a function of the lower cut applied on the dielectron invariant mass. 
The solid lines represent the pure DY background while the dashed ones include also the PI contribution. 
Different colors refer to different stages of the LHC as described in the legend. 
Standard acceptance cuts are applied ($|\eta_l| < 2.5$ and $p_T^l > 20~$GeV) as well as the declared efficiency of the electron channel \cite{Khachatryan:2014fba}. NNLO QCD corrections are accounted for in the DY term \cite{Hamberg:1990np}.}
\protect{\label{fig:DY_PI}}
\end{figure}

\noindent
In Fig. \ref{fig:13vs8}a, we plot the differential cross section in the dilepton invariant mass for the PI process at the 8 TeV LHC (blue line) and the 13 TeV LHC (red line). In Fig. \ref{fig:13vs8}b, we show the ratio of these two spectra (magenta line) together with the analogous ratio for the DY process (green line), that is the production of lepton pairs induced by a quark-antiquark interaction via the exchange of off-shell SM $\gamma$ and $Z$ gauge bosons. As one can see, there is an increase of the PI differential cross section with the collider energy, even if less substantial than in the DY case. For an invariant mass $M_{ll}=3$ TeV, the PI differential cross section has increased roughly by a factor of five, in the upgrade from RunI to RunII, while the DY one has gone up by a factor of twenty. 

\noindent
The importance of the PI contribution seems to be reduced with the energy upgrade of RunII and this effect is mainly due to the stronger relative enhancement of the quark PDFs in comparison with the QED PDFs. This is indeed the result displayed in Fig. \ref{fig:DY_PI}a. There, we show the complete result, given by the sum of PI and DY contributions to the differential cross section, normalized to the DY result. The relative importance of the PI contribution with respect to the DY differential cross section appears in principle much bigger at the 8 TeV LHC. However, one needs to consider the available integrated luminosity before drawing conclusions. The role of the luminosity is displayed in Fig. \ref{fig:DY_PI}b where we show the number of events in the dielectron channel, which are expected at each of the three representative stages of the LHC described in the legend. The solid lines show the DY cumulative cross section, computed from the dielectron invariant mass threshold on, while the dashed ones include also the PI contribution to the same final state.
NNLO corrections to the DY term, obtained using the WZPROD code at different invariant mass points\footnote{We are aware of the fact that the NNLO $k$-factor obtained this way (i.e., fully inclusively) is not fully justified for a calculation that involves cuts on the phase space. However, we emphasise that the actual central value of the DY prediction (or its error, especially in relation to the PI case) is not our main concern here and that this practise is often used in experimental analyses. Conversely, in our extraction of LHC sensitivities to DY, estimates of higher order effects are necessary.} \cite{Hamberg:1990np, Harlander:2002wh}, are included, while for the PI process, on the other hand, we are considering only the LO calculation in QED. In the calculations of the expected number of events and of the significances of BSM signals we also include the overall efficiency factor given in Ref. \cite{Khachatryan:2014fba}.

\noindent
As one can see from Fig. \ref{fig:DY_PI}b, with the luminosity collected during RunI ($L\simeq 20~fb^{-1}$) the number of dilepton events predicted within the SM goes to zero for invariant masses beyond $M_{ll}=1.8$ TeV. Up to these energy scales, the PI contribution is at most 25\% of the standard DY cross section. The extracted exclusion bounds on the mass of an hypothetical narrow-width $Z^\prime$-boson have been set in a 2.5--3.2 TeV range, depending on the specific theoretical model. This range lies in a dilepton invariant mass window where the SM background is zero. One can thus conclude that the PI effect was not decisive for the $Z^\prime$-boson searches in the dilepton channel performed at the LHC RunI and for the exclusion bounds that have been from there derived.

\begin{figure}[t]
\centering
\includegraphics[width=0.45\linewidth]{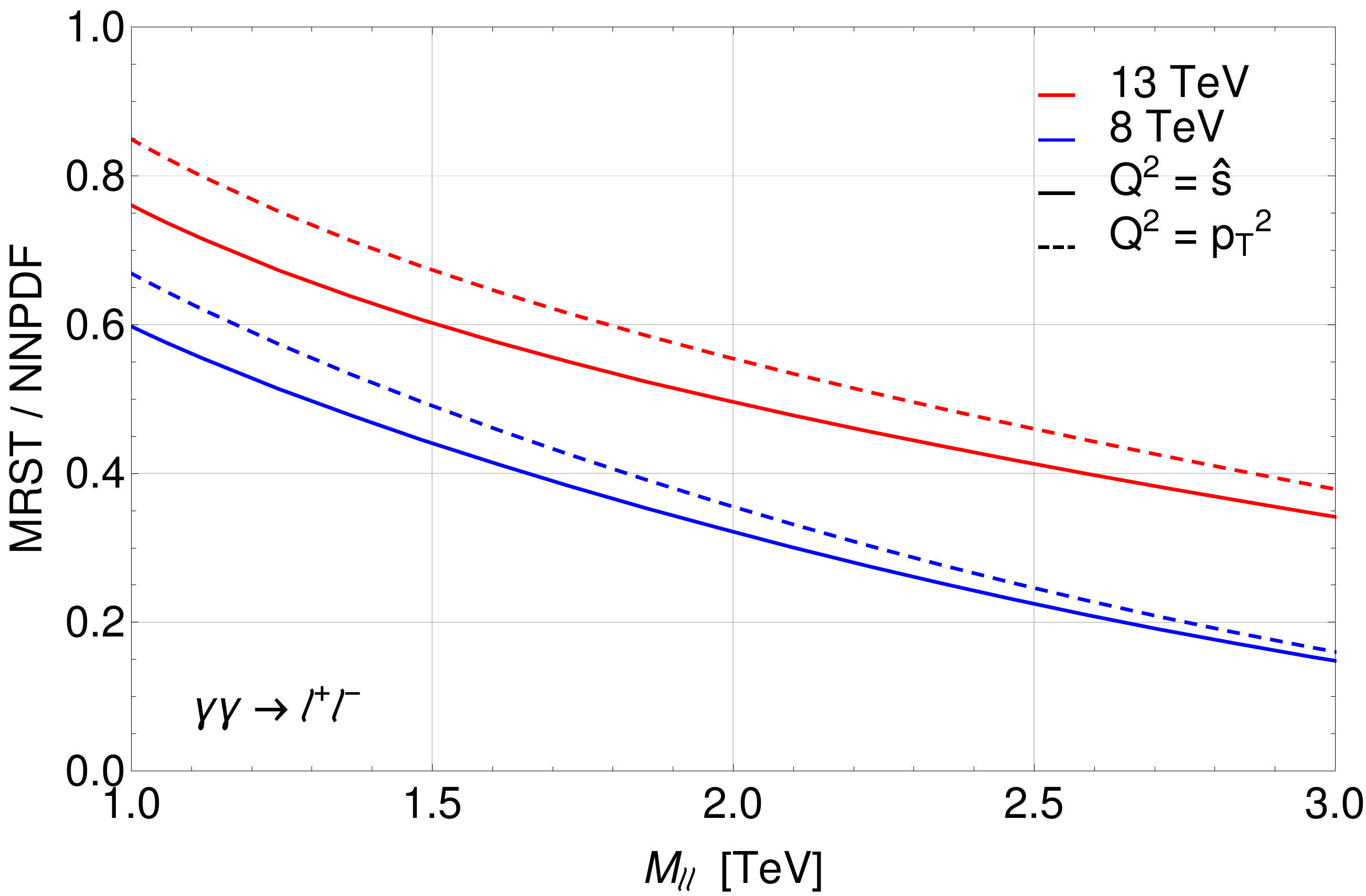}{(a)}
\includegraphics[width=0.45\linewidth]{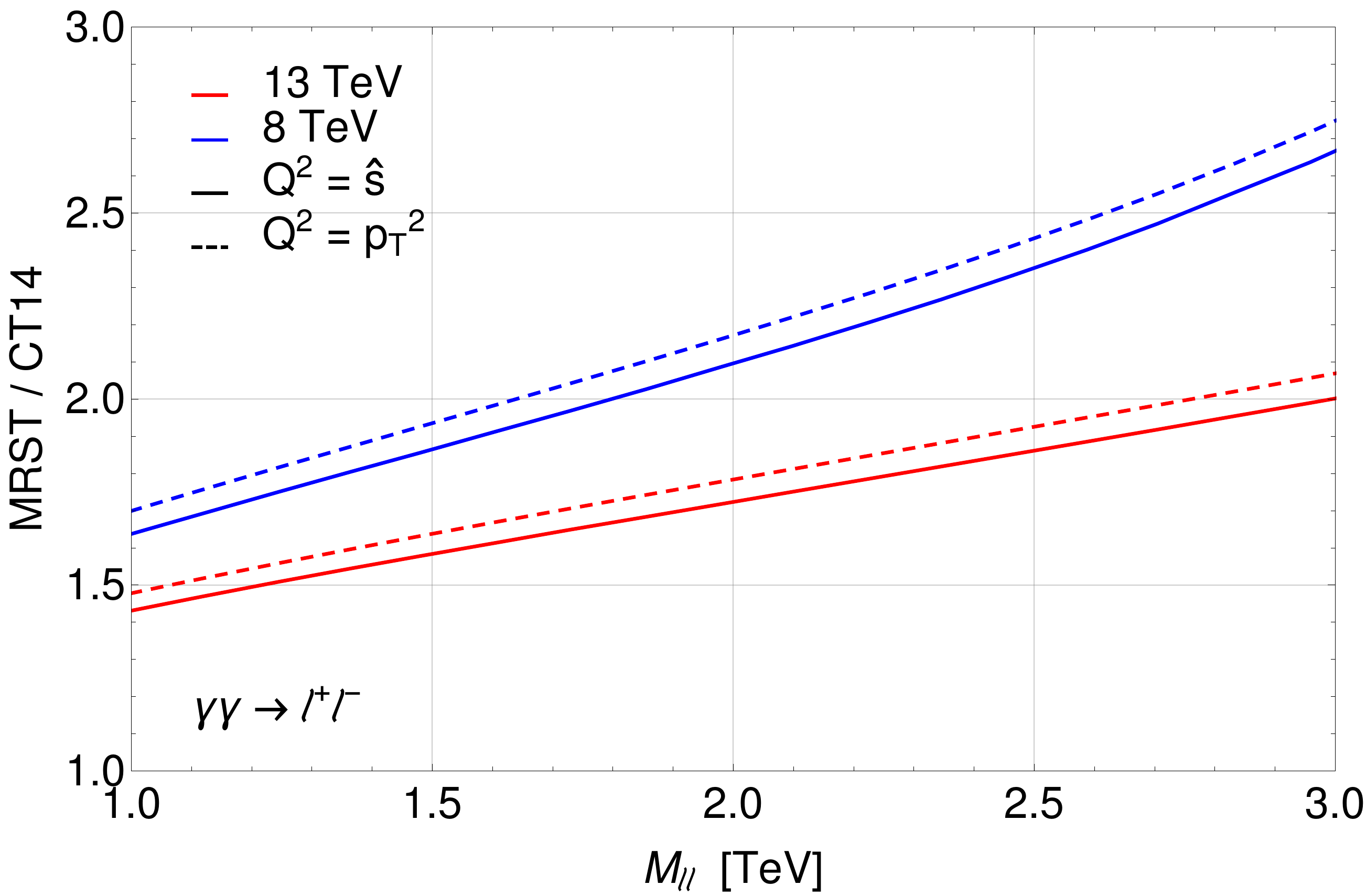}{(b)}
\includegraphics[width=0.45\linewidth]{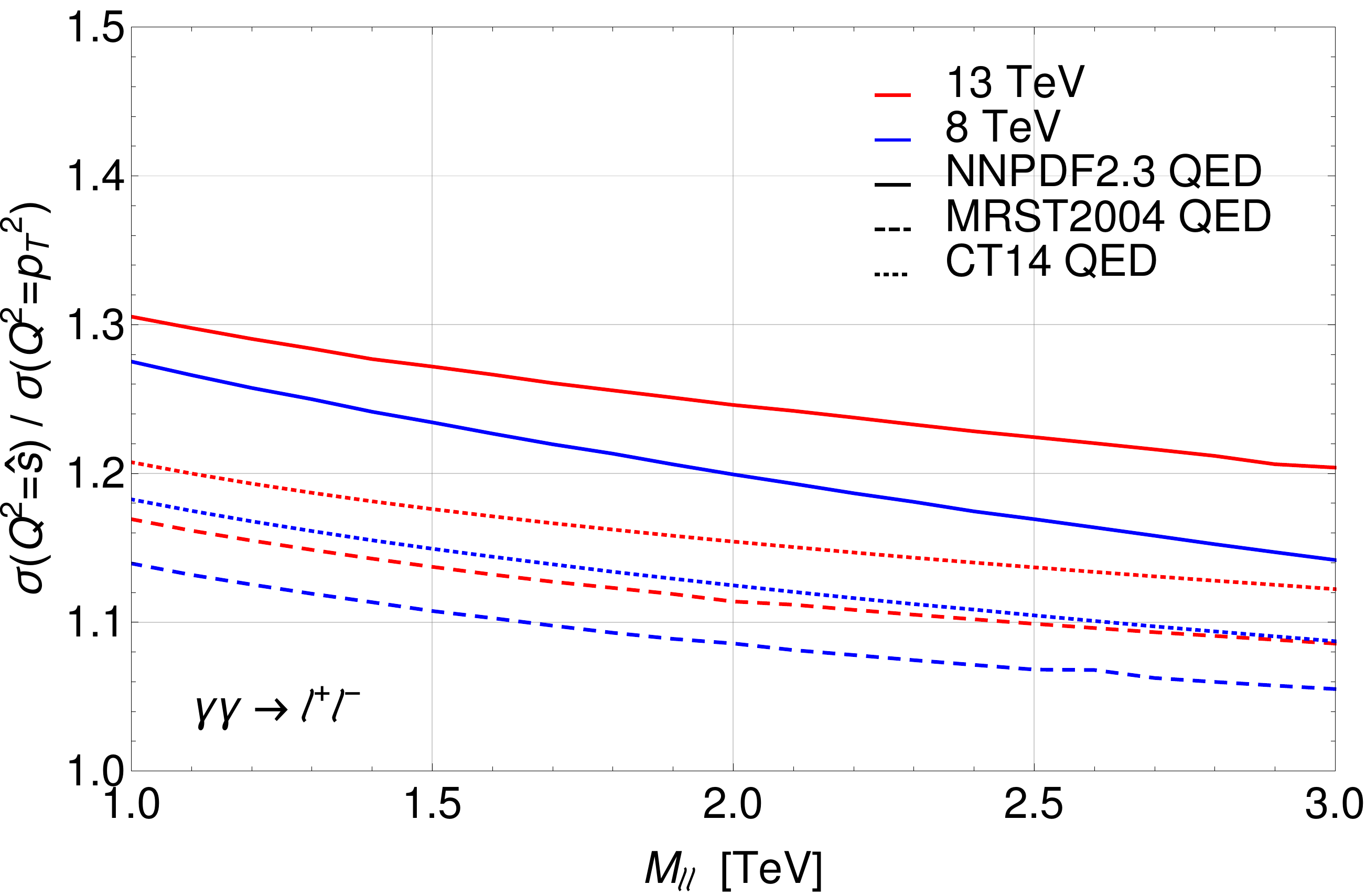}{(c)}
\includegraphics[width=0.45\linewidth]{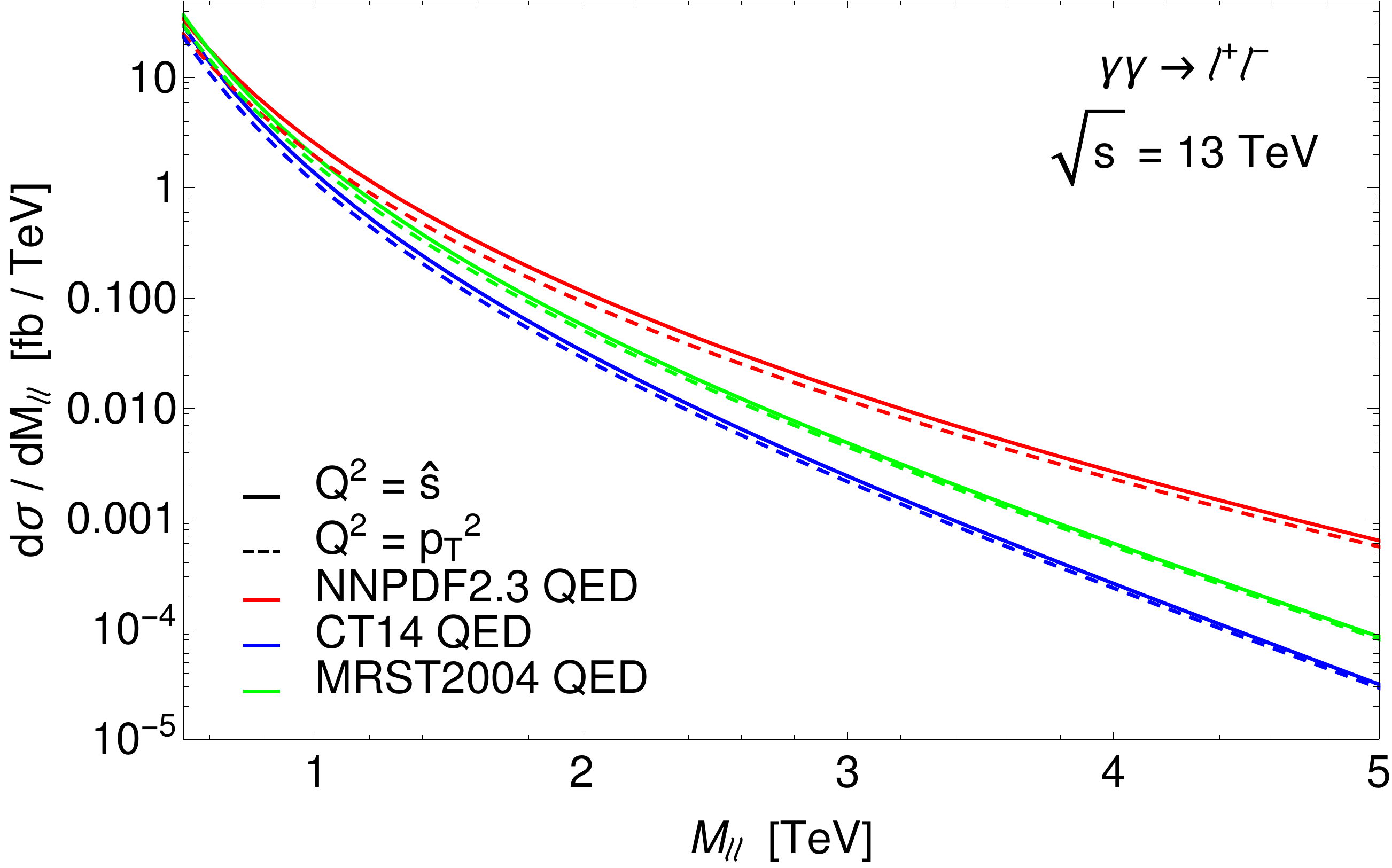}{(d)}
\caption{(a) Ratio between the central values of the photon induced dilepton spectrum obtained with the MRST2004QED and the NNPDF2.3QED sets. 
(b) Same as (a) but now comparing the MRST2004QED and the CT14QED sets. 
(c) Ratio between the dilepton spectrum at two different factorisation scales, $Q^2=\hat s$ and $Q^2=P_T^2$, for MRST2004QED (dashed lines), for CT14QED (dotted lines) and NNPDF2.3QED (solid lines) at the 8 TeV LHC (blue lines) and the 13 TeV LHC (red lines).
(d) Differential cross section distribution for the PI process as predicted by the three PDF collaborations specified.
Standard acceptance cuts are applied ($|\eta_l| < 2.5$ and $p_T^l > 20~$GeV).}
\protect{\label{fig:PDFsCompare}}
\end{figure}

\noindent
Different is the situation at the present LHC RunII. In one year time, the collected luminosity is expected to be $L\simeq 30~fb^{-1}$. The number of SM dilepton events will be different from zero up to invariant masses of the order of 2.5 TeV, that is at the edge of the present exclusion bounds on the mass of possible narrow-width extra $Z^\prime$-bosons. In this instance, the PI contribution could impact on the $Z^\prime$-searches performed by the experimental collaborations. A default procedure, applied for example by the CMS collaboration in $Z^\prime$-searches, is to simulate the SM background with a functional form whose parameters are fitted to the predictions obtained by Monte Carlo (MC) simulations. The expected SM background given by the functional form is then normalized to the data in a dilepton invariant mass window on the left-hand side of the hypothetical $Z^\prime$-boson pole mass. 
The lower edge of this window is determined by the requirement of collecting 400 events.

\begin{figure}[t]
\centering
\includegraphics[width=0.45\linewidth]{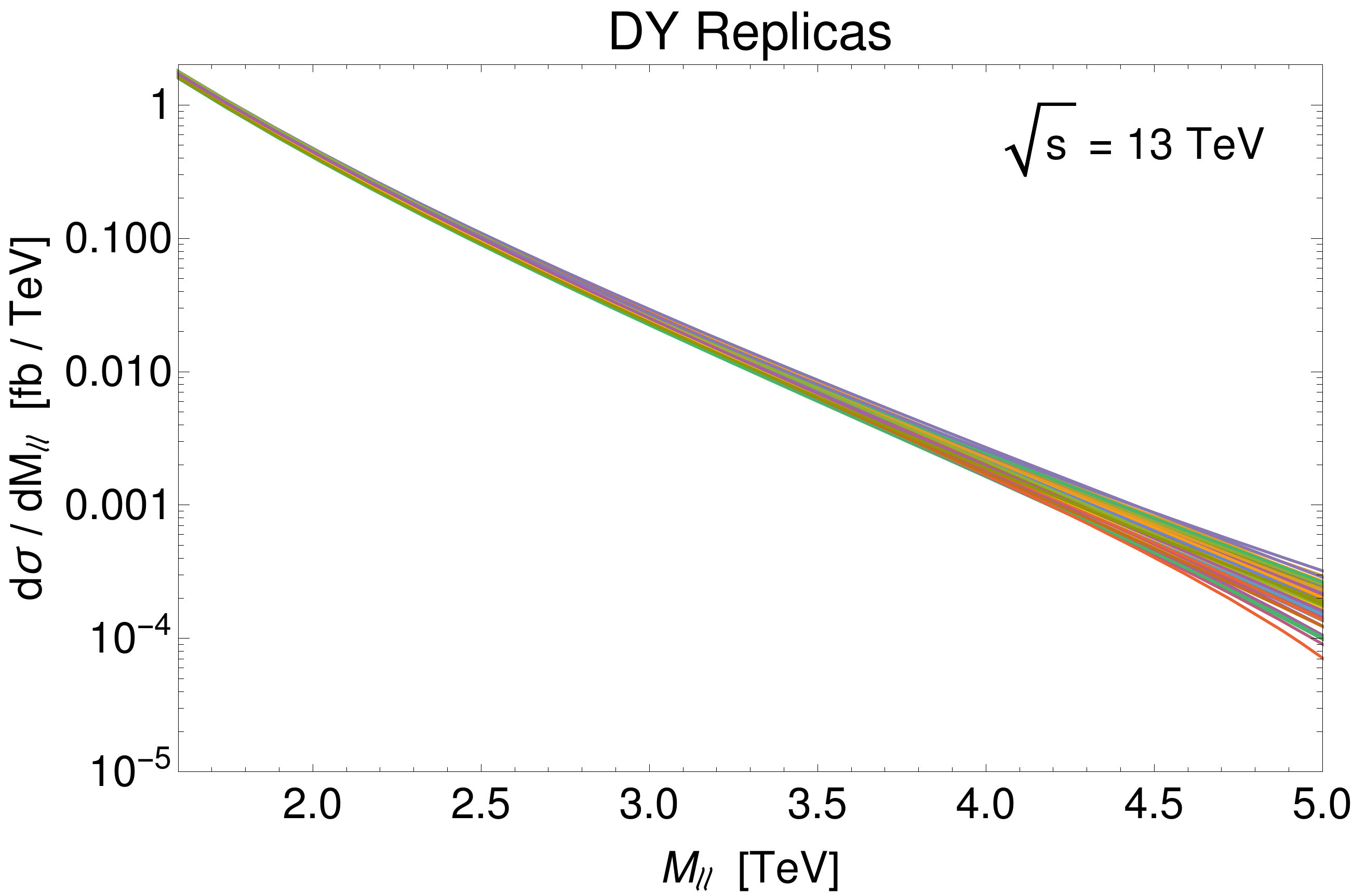}{(a)}
\includegraphics[width=0.45\linewidth]{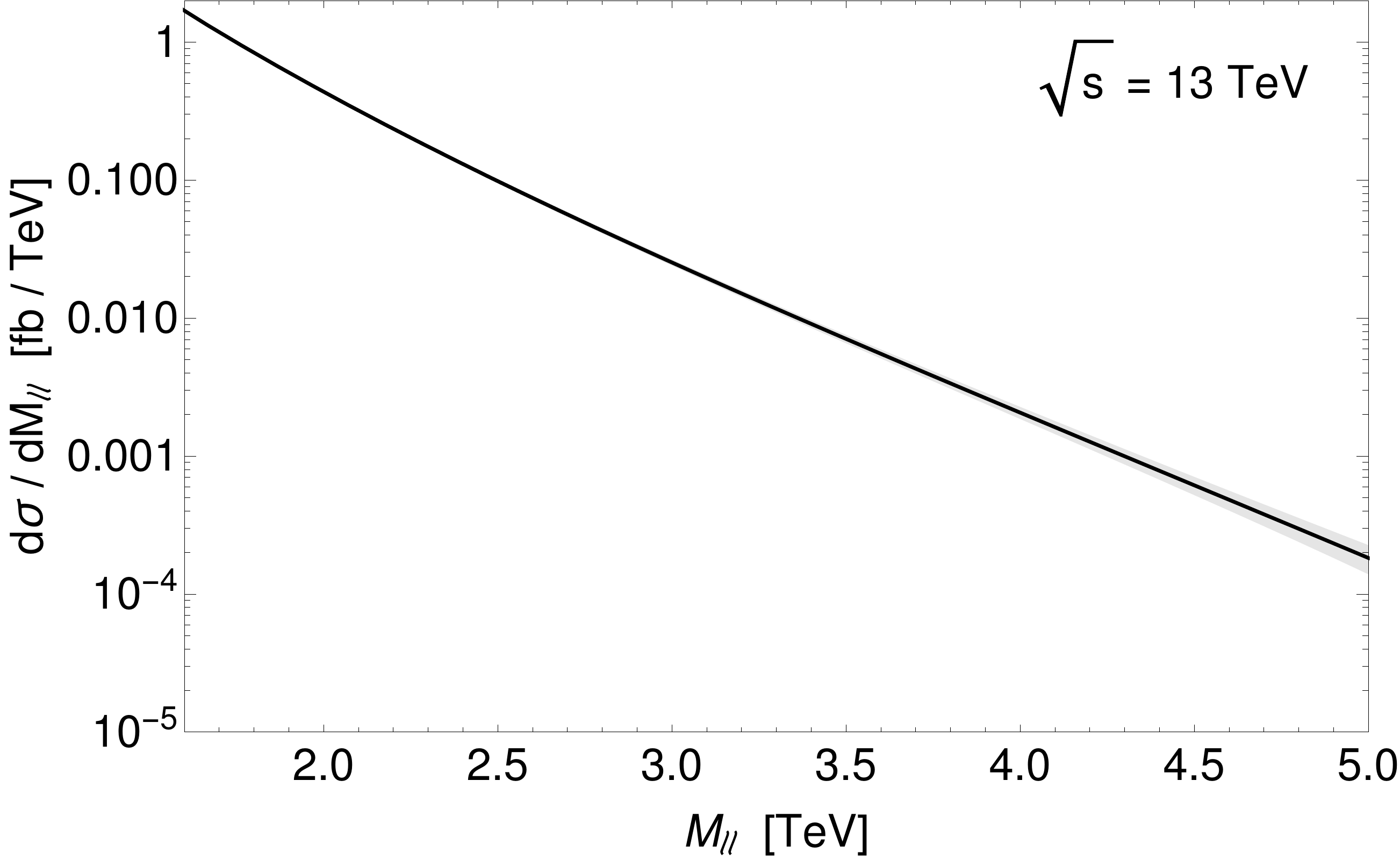}{(b)}
\includegraphics[width=0.45\linewidth]{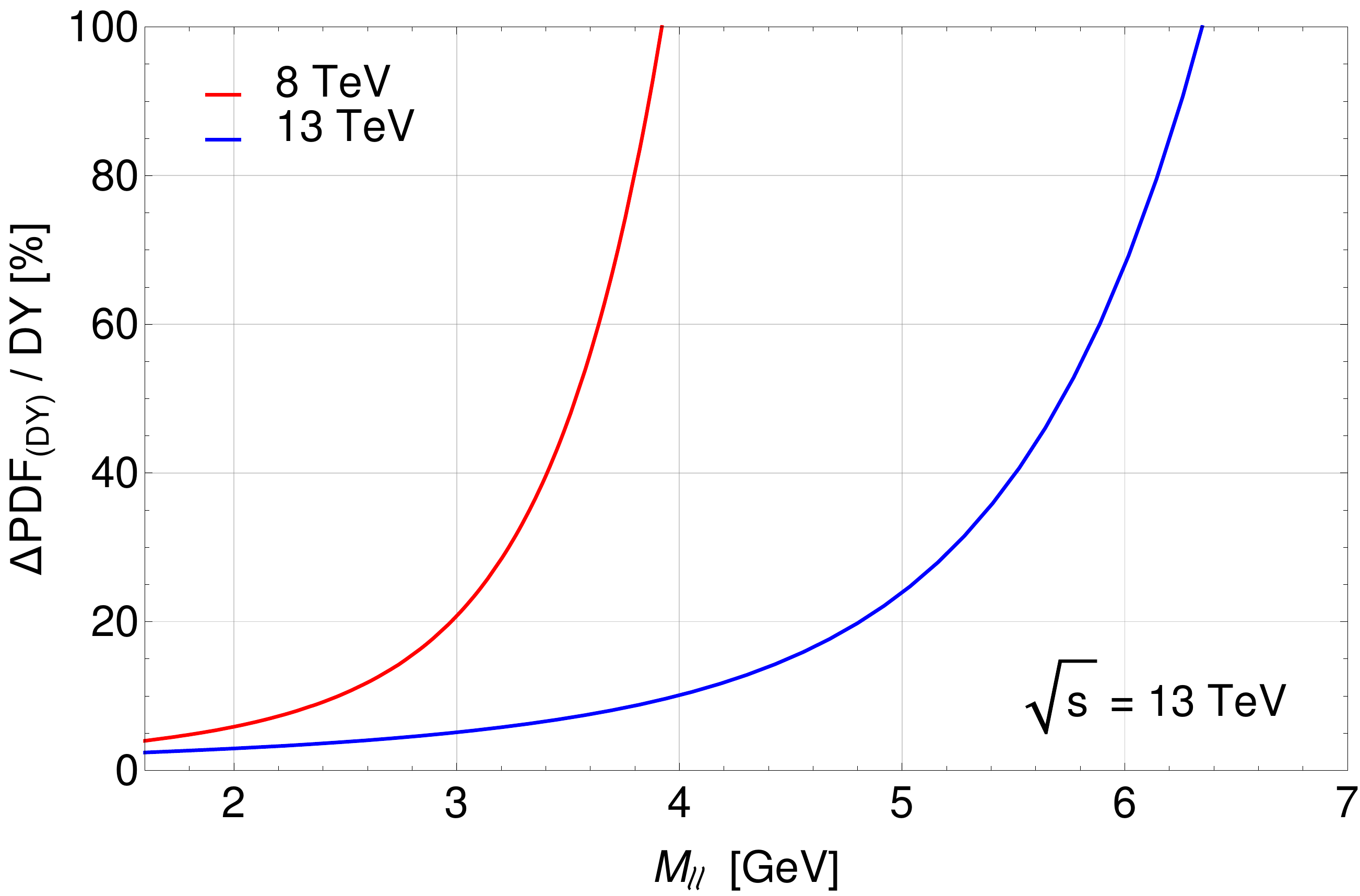}{(c)}
\caption{(a) DY dilepton spectrum for the 100 NNPDF replicas. 
(b) Central value and PDF error for the DY dilepton spectrum via NNPDF.
(c) Relative size of the PDF error for the DY process at the 8 TeV (blue line) and 13 TeV (red line) LHC. Standard acceptance cuts are applied ($|\eta_l| < 2.5$ and $p_T^l > 20~$GeV).}
\protect{\label{fig:DY_replicas}}
\end{figure}

\noindent
If we consider the present lower limit from which new searches are starting, $M_{Z'}=2.5$ TeV, and take a mass range on its left-hand side such to satisfy this condition, we find that the PI contribution at the 13 TeV LHC can be of the order of 25\% of the DY differential cross section usually accounted for. The relative magnitude of the PI versus DY cross section is the same as during RunI. What changes is the number of expected SM background events, in the search window where the unbinned likelihood fitting procedure is performed, which are near the hypothetical $Z^\prime$-boson mass. At RunI, this number was null above $M_{ll}=1.8$ TeV, at RunII is not anymore. In this respect, the PI contribution can start playing a sizeable role at present searches. Its importance will increase in the future, when the project luminosity $L\simeq 300~fb^{-1}$ will be reached. In that case, the number of SM background events is estimated to be different from zero up to about 3.5 TeV. At these mass scales, the PI contribution goes up to roughly 60\% of the DY cross section. It is thus advisable to take the PI contribution into account when simulating the SM background to the dilepton final states, for any sensible interpretation of the data.

\noindent
Such a PI background is surely important for BSM searches of narrow-width extra $Z^\prime$-bosons, as we discussed up to now, for it could affect the extracted $Z^\prime$ mass bounds in regions of the dilepton spectrum where the number of expected SM events is different from zero. 
However, it becomes definitely an issue for non-resonant searches in the same channel, like for contact interactions. In this case, a BSM signal would appear like either an excess of events evenly spread over the SM background or as a plateau. The experimental search in this case relies on a counting strategy. The theoretical interpretation of such a non-shaped signal requires a particularly accurate knowledge of the SM background. To this end, the computation of the central value of the PI cross section and the determination of its theoretical uncertainty is mandatory for any reliable interpretation of the experimental results. We discuss in more detail the impact of the PI background on BSM searches for both examples of new physics signals in Sect.~\ref{sec:BSM}.

\noindent
A recent analysis by ATLAS at 8 TeV LHC, published in Ref. \cite{Aad:2014wca}, is dedicated to non-resonant searches in the dilepton channel. There, the PI background is explicitly included. 
It is generated via Pythia8, adopting the MRST2004QED PDFs \cite{Martin:2004dh}. According to the plots on the dilepton spectrum, shown there, the PI contribution is marginal compared to the DY one. This is not consistent with what we find with the NNPDF2.3 set in the same dilepton invariant mass region. In order to understand this discrepancy, in Fig. \ref{fig:PDFsCompare}a we compare the two PI distributions in the dilepton invariant mass obtained with the two different sets of PDFs. We consider only the central values, neglecting the PDF uncertainty as that is not available for the MRST2004QED set. To visualize the difference between the two results, we plot the ratio between the dilepton spectrum computed with the MRST2004QED PDF and the spectrum evaluated with the NNPDF2.3QED set. We consider the two collider energies: 8 TeV (blue line) and 13 TeV (red line). In addition, we take two possible factorization scales $Q^2=\hat{s}$ (solid lines) and $Q^2=P_T^2$ (dashed lines), in order to analyse the scale choice dependence. 

\noindent
As one can see in Fig. \ref{fig:PDFsCompare}a, the central value of the PI contribution obtained with NNPDF is significantly larger than that obtained with MRST photon PDFs. The difference increases with the mass scale. In Fig. \ref{fig:PDFsCompare}b we have repeated the same exercise for the CT14QED set (more precisely we use the CT14QED\_inc set which includes also the elastic scattering component) in comparison with the MRST set and the difference we have found is also sizeable, with the central value predicted by the CT14 collaboration being smaller than the MRST one. Changing the factorization scale does not affect much the conclusion. It adds in fact only a few tens of percent uncertainty. At the 8 TeV LHC, that is the energy at which the ATLAS analysis was carried out, the MRST result is on average a factor of 2--3 smaller than the NNPDF one in the explored search window. This is the origin of the difference we are observing. 
In Fig. \ref{fig:PDFsCompare}c, we analyse the dependence on the factorization scale in more detail. We plot the ratio between the dilepton spectrum evaluated at two different factorization scales, $Q^2=\hat{s}$ and $Q^2=P_T^2$. 
We computed the three QED PDF sets predictions at the 8 TeV and 13 TeV LHC. 
The result is that the scale dependence is more pronounced for the NNPDF. It ranges 
between 14\%--30\% in the shown mass window, while it varies between 6\%--18\% when using MRST and between 9\%--20\% with the CT14 set.
In Fig. \ref{fig:PDFsCompare}d we plot the central value for the PI process obtained with the three PDF sets. Clearly the differences between the predictions from each set are substantial. 

\noindent
So far, we have discussed the central value of the PI contribution. The outcome is that in the mass range of interest at the present LHC RunII, $M_{ll}\ge 2.5$ TeV, the PI contribution to the dilepton spectrum can range between 30\%--200\% of the DY cross section and increases with the mass scale. The PI (differential) cross section is definitely sizeable and comparable in magnitude to the DY background. Moreover, it falls less rapidly than the DY contribution with increasing dilepton invariant mass. As a consequence, not only the magnitude but also the shape of the overall SM dilepton spectrum can change, owing to the PI contribution. The second message to extract is that the difference between the three PDF collaborations predictions for the PI cross section is significant and increases with the mass scale. 
The NNPDF central value is larger than the MRST one by roughly a factor of 2--3 in the region of interest, and roughly a factor of 3--6 larger than the CT14 best fit in the same region.
This is already a measure of the uncertainty on the PI central value. In the next section, we shall address the proper QED PDF uncertainty, by computing the NNPDF error. We shall see that the difference between the three central values of the photon induced dilepton cross section is completely included in the PDF error band estimated via the NNPDF2.3QED set. 
Similar results have been found by the authors of Ref. \cite{Ball:2013hta}.

\section{Replicas method for the PDF uncertainty}\label{sec:PDFerror}

\begin{figure}[t]
\centering
\includegraphics[width=0.45\linewidth]{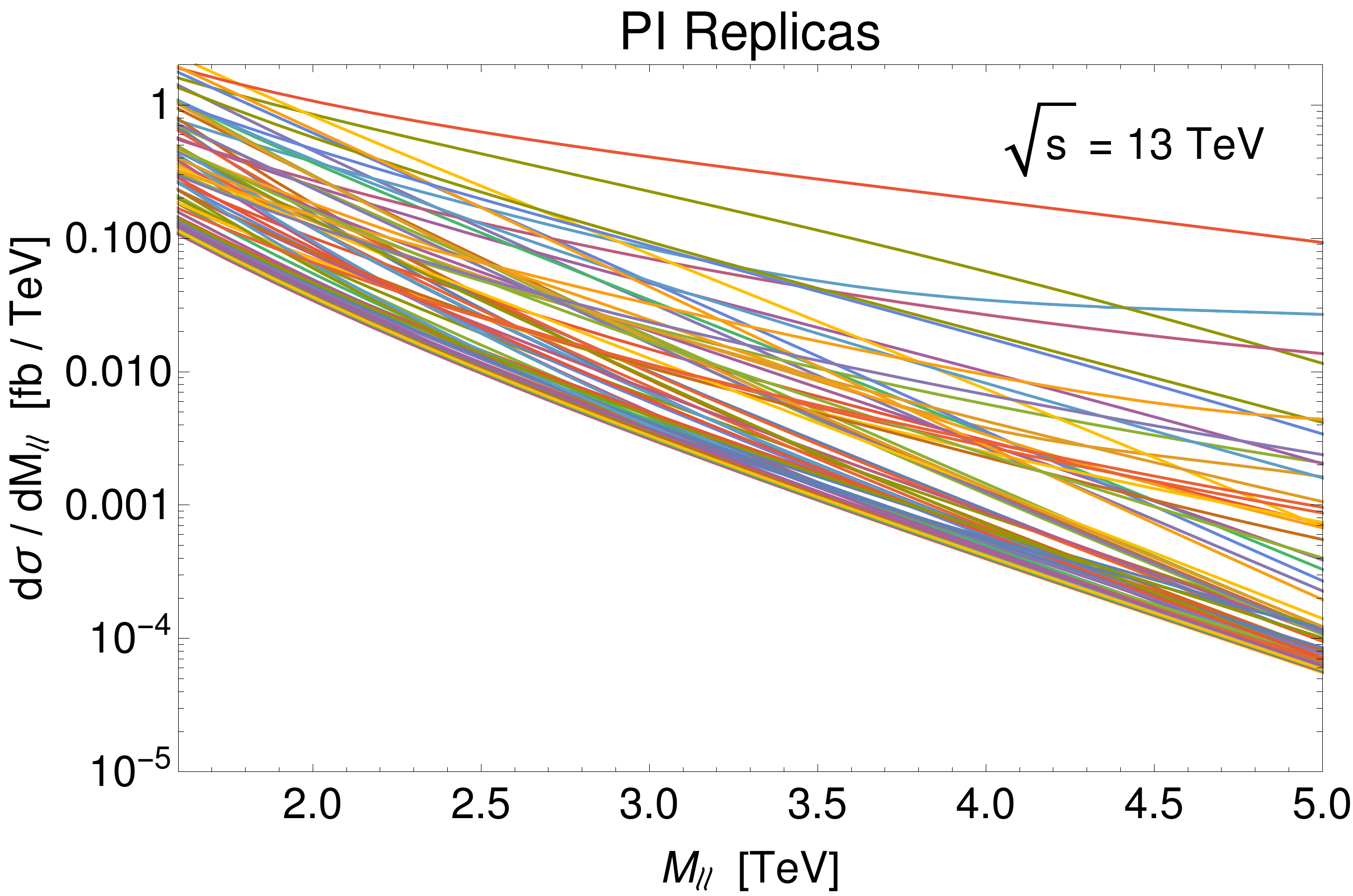}{(a)}
\includegraphics[width=0.45\linewidth]{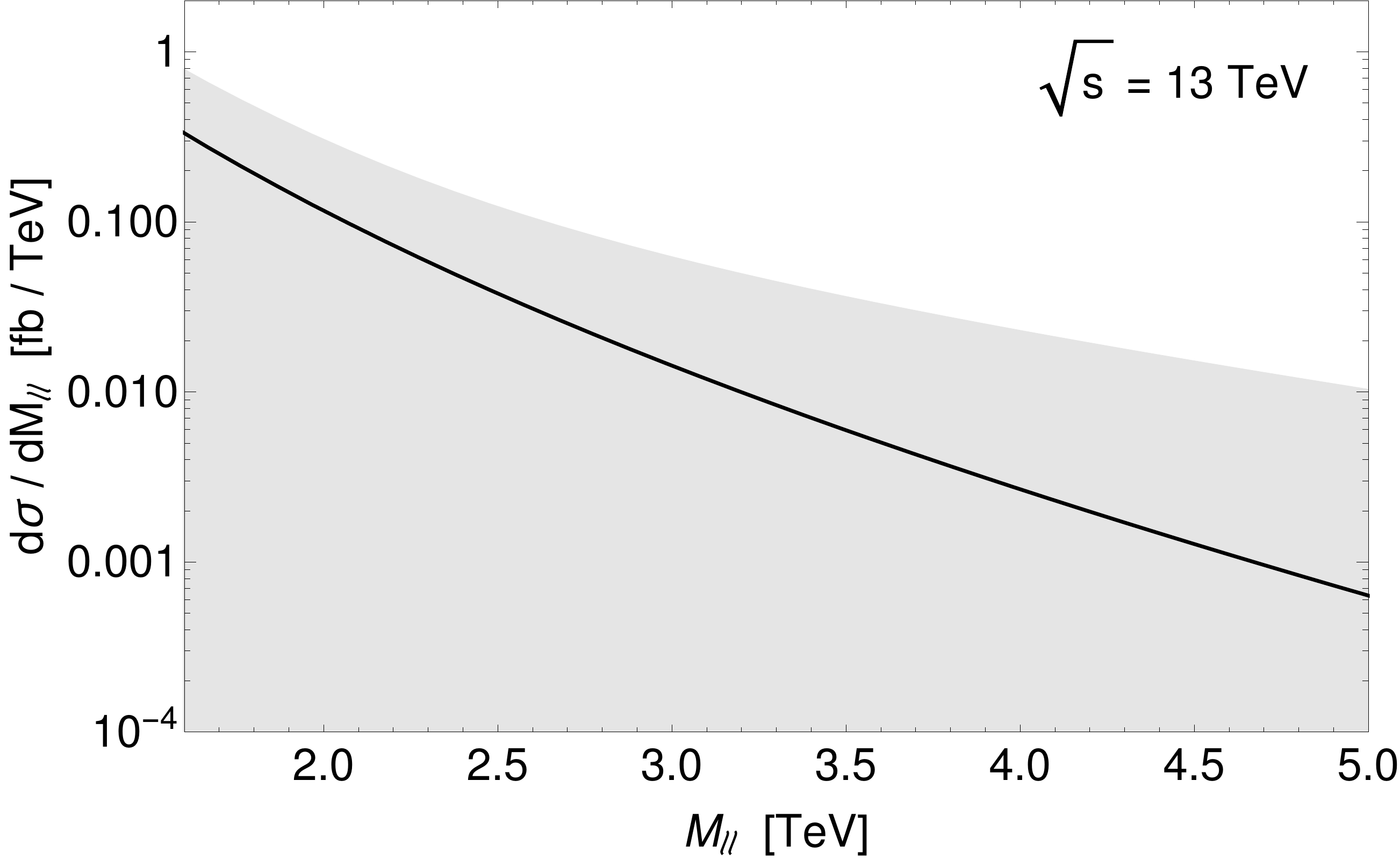}{(b)}
\includegraphics[width=0.45\linewidth]{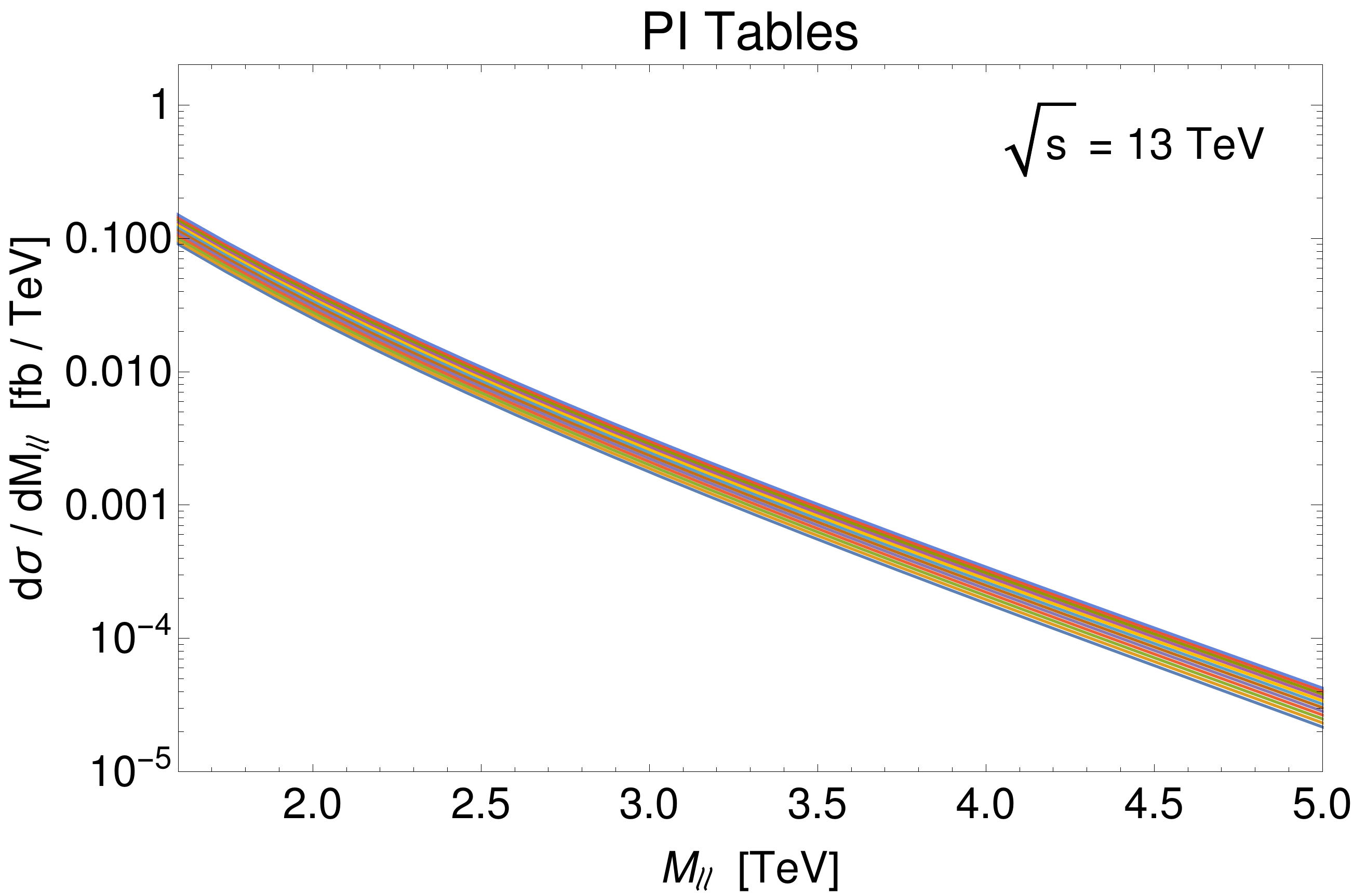}{(c)}
\includegraphics[width=0.45\linewidth]{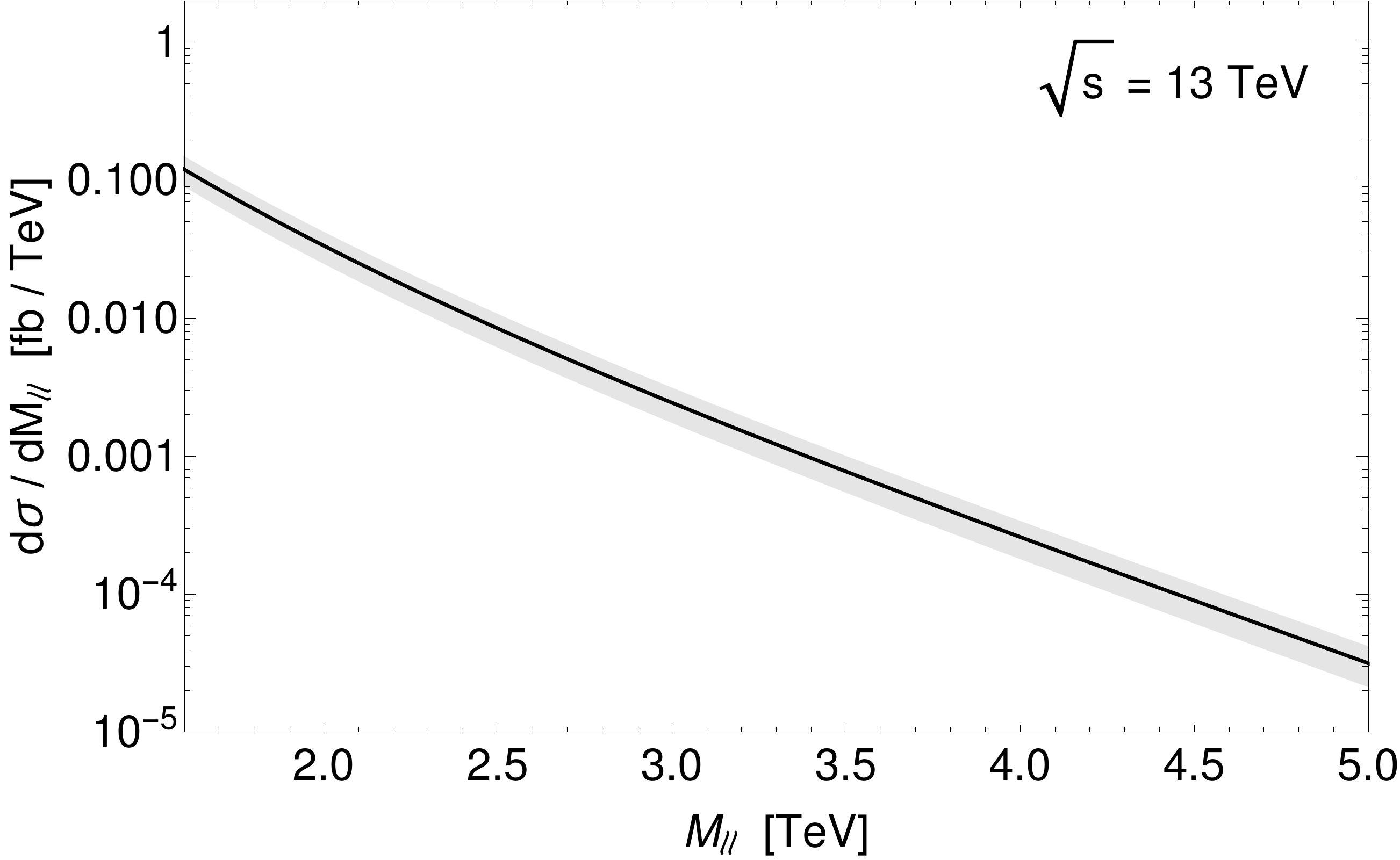}{(d)}
\caption{(a) PI dilepton spectrum for the 100 NNPDF replicas. 
(b) Central value and PDF error for the PI dilepton spectrum in NNPDF.
(c) PI dilepton spectrum for the first 15 CT14 tables, which parametrise an initial momentum fraction of the proton total momentum carried by the photon ranging from 0.00\% to 0.11\%.
(d) Central value and PDF error for the PI dilepton spectrum extracted from CT14 tables and upper limits on the momentum fraction as quoted in \cite{Schmidt:2015zda}.
Standard acceptance cuts are applied $|\eta_l| < 2.5$ and $p_T^l > 20~$GeV.}
\protect{\label{fig:PI_replicas}}
\end{figure}

In this section, we estimate the theoretical error on the double dissociative PI dilepton production coming from the PDF uncertainty. This can be considered as a systematic error on the determination of the dilepton spectrum. The various quark (antiquark) PDFs from different collaborations come with specific procedures to estimate the error on the central value. 
The various PDF collaboration groups decided to follow different approaches. CTEQ and MRST apply the Hessian method that exploits PDF eigenvalues \cite{Alekhin:2011sk,Accomando:2015cfa}. In this approach, the error is estimated from the standard deviation of a limited number of central values coming from the difference of paired PDF fits (order 20 pair of fits). 
In this work we mainly followed the replicas method adopted by the NNPDF collaboration. The error on the PDF central value is computed as the standard deviation of a large set of replicas (order 100) that represent other possible fits of the experimental data \cite{Ball:2013hta,Ball:2011gg}. For any observable, the central value is defined as the average of the different replicas and its error is given by the standard deviation as summarized by the following equations, 
\begin{equation}
O_0 = \langle O\rangle = \frac{1}{N}\sum_{k=1}^N O_k,
\end{equation}
\begin{equation}
\left(\Delta O\right)^2 = \frac{1}{N}\sum_{k=1}^N (O_k - O_0)^2,
\end{equation}
where $O_k$ ($k=1,...,N$) are the $N$ replicas. Following this approach, we have evaluated the differential cross section for the hundred NNPDF replicas for both the DY and PI processes. 
The good quality of the quark (antiquark) fit translates into a rather satisfactory prediction for the DY dilepton spectrum. This is shown in Fig. \ref{fig:DY_replicas}a where we plot the dilepton invariant mass distribution for all the replicas. The result of the averaging procedure gives the central value and the error band visible in Fig. \ref{fig:DY_replicas}b. 

\noindent
At the LHC RunII with 13 TeV, the PDF uncertainty coming from the large-$x$ region is pushed towards higher dilepton invariant masses, compared to RunI. More in detail, the relative PDF error grows above 10\% for $M_{ll}\ge 4$ TeV and goes up sharply to 80\% at the LHC potential edge around $M_{ll}\simeq 6$ TeV, as shown in Fig.\ref{fig:DY_replicas}c.
\begin{figure}[t]
\centering
\includegraphics[width=0.45\linewidth]{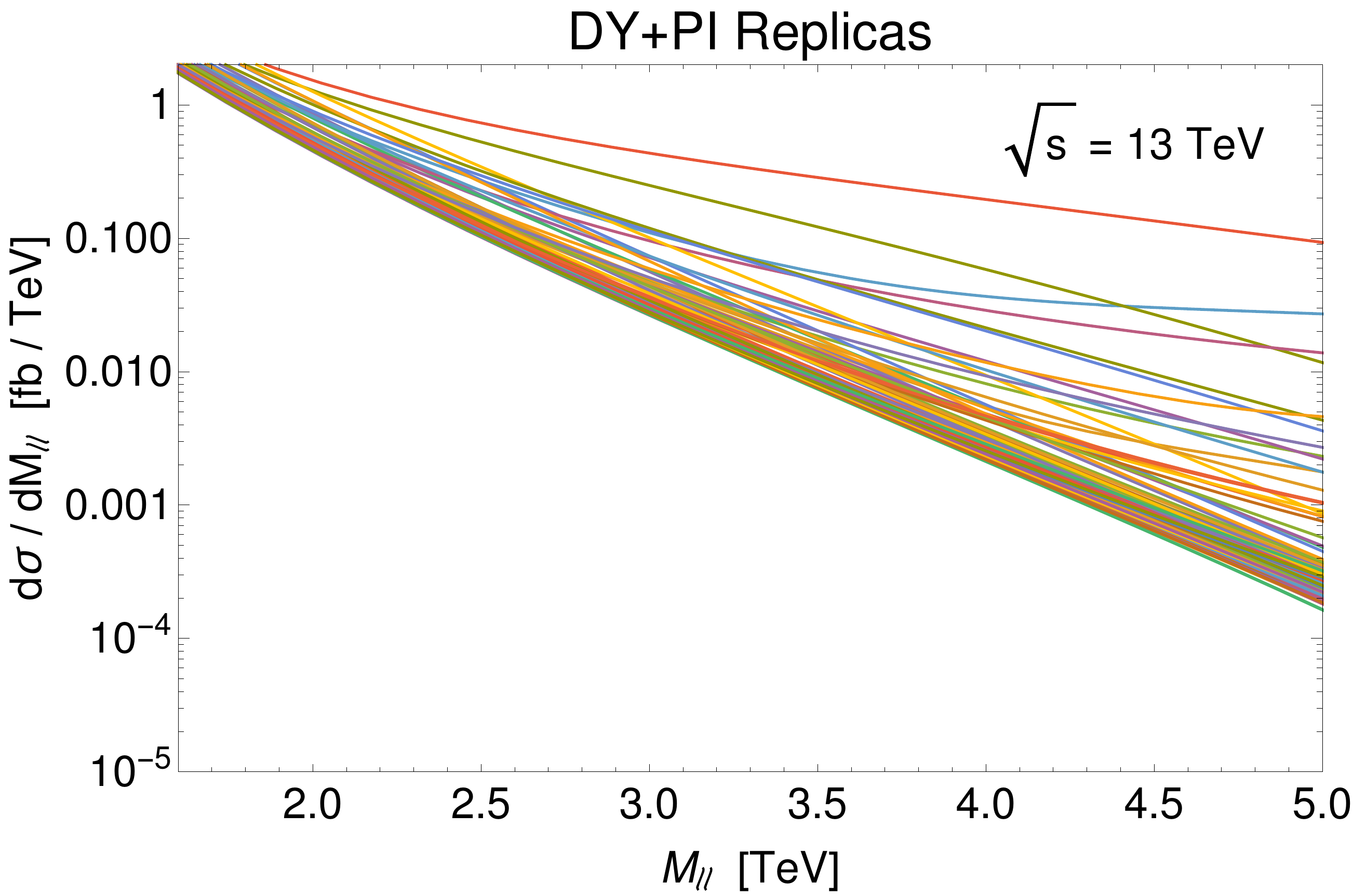}{(a)}
\includegraphics[width=0.45\linewidth]{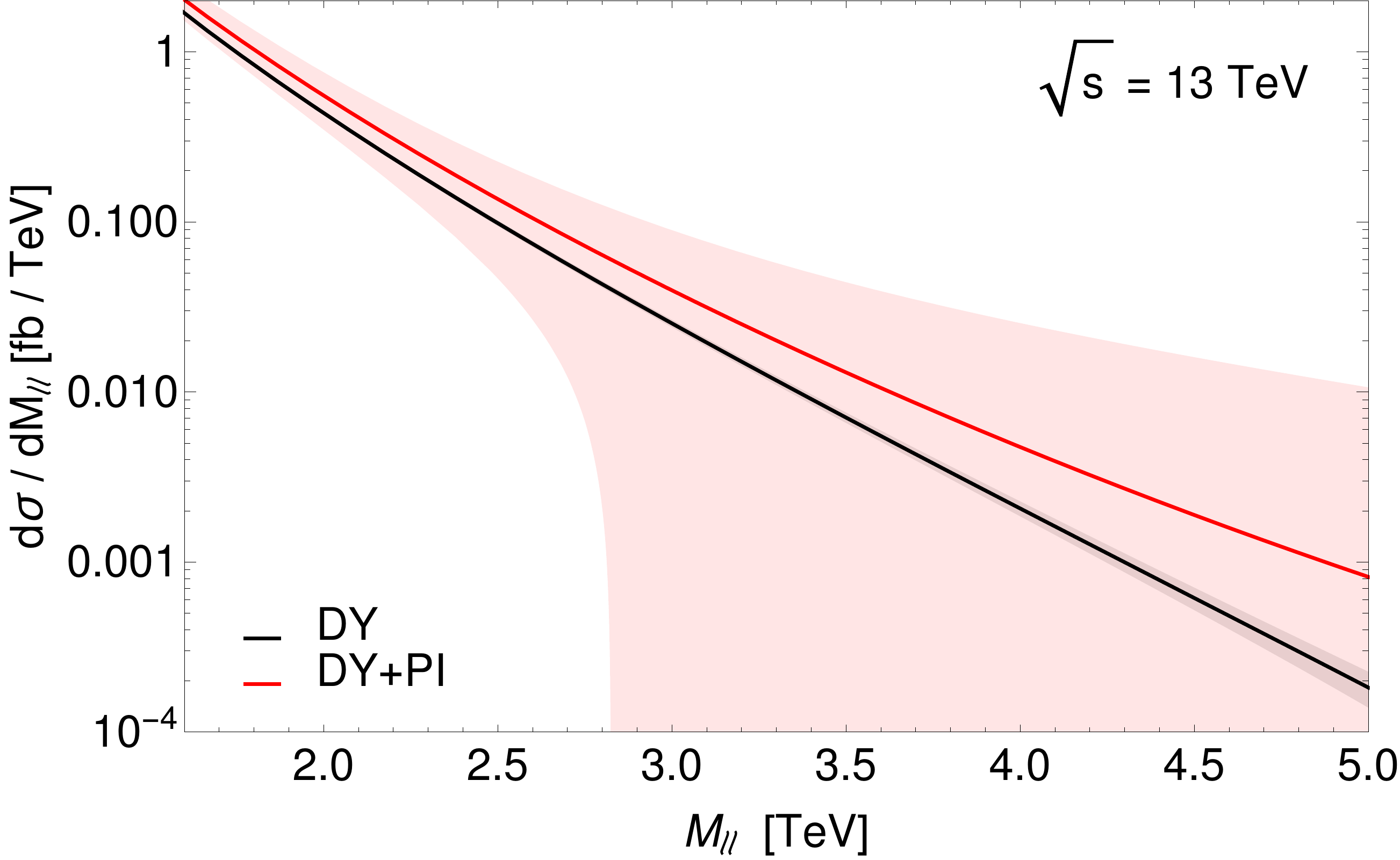}{(b)}
\includegraphics[width=0.45\linewidth]{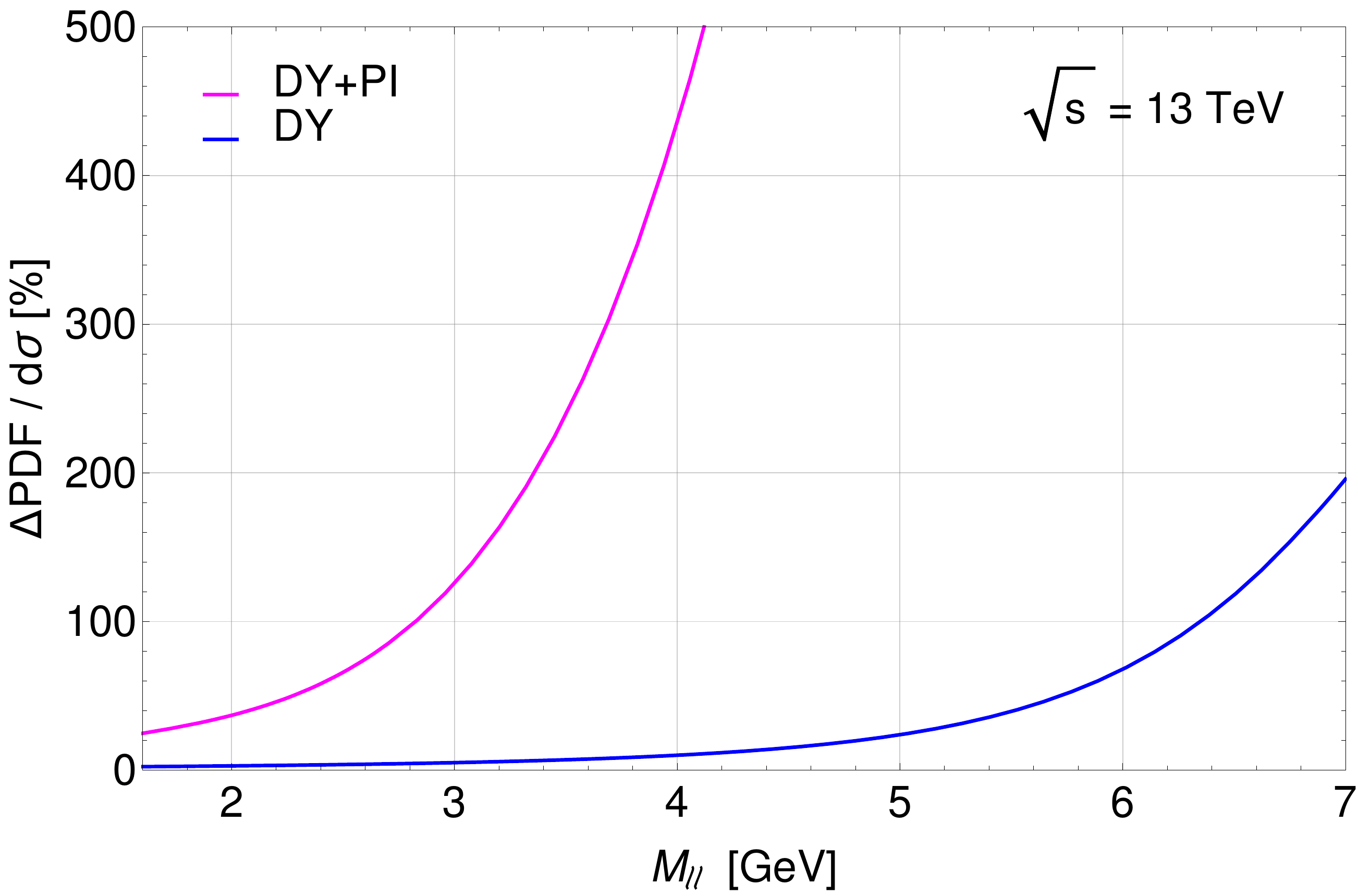}{(c)}
\caption{(a) Prediction of the DY+PI dilepton spectrum for the 100 NNPDF replicas. 
(b) central value for the DY (black line) and DY+PI (red line) dilepton spectrum from NNPDF including the PDF error band for the two cases.
(c) Relative impact of the PDF uncertainties with (magenta line) and without (blue line) the PI contribution. Standard acceptance cuts are applied ($|\eta_l| < 2.5$ and $p_T^l > 20~$GeV).}
\protect{\label{fig:DY_PI_replicas}}
\end{figure}
The theoretical error on the DY process initiated by a quark-antiquark interaction looks reasonably under control over a large portion of the dilepton spectrum. Unfortunately, the situation is much worse for the PI process. The uncertainty on the QED PDF results in more scattered predictions, as shown in Fig. \ref{fig:PI_replicas}a where we plot the PI dilepton spectrum for the hundred different replicas in NNPDF2.3QED. The central value and its error, estimated as before, are shown in Fig.\ref{fig:PI_replicas}b. 

An estimation of the photon PDF uncertainty is also given in the CT14QED set, which is provided with 31 PDFs tables, each one parameterising the photon PDF under the constraint that the photon is carrying a fixed fraction of the proton total momentum.
Thus we have predictions for momentum fraction carried by the photon growing from 0.00\% to 0.30\%. It is found \cite{Schmidt:2015zda} that at 90\% C.L. the maximum momentum fraction allowed for the photon is 0.14\%, and similar analysis give a maximum of 0.11\% at 68\% C.L.. 
Following this result, we have extracted the CT14QED central value prediction averaging the results obtained using the appropriate 12 tables, while the 68\% C.L. area is twice the 1$\sigma$ deviation band. The prediction from the CT14 first 12 tables is shown in Fig. \ref{fig:PI_replicas}c, while its central value and 1$\sigma$ error band are shown in Fig. \ref{fig:PI_replicas}d. 
Not only do the NNPDF2.3QED and CT14QED central values differ significantly, as already seen in the previous section, but also the estimates of the photon PDF uncertainty are very different between NNPDF2.3QED and CT14QED. 

Once the DY and PI contributions to the dilepton final state are added up, the overall scenario becomes more difficult to handle. 
The dilepton spectrum for the different NNPDF2.3QED replicas is shown in Fig. \ref{fig:DY_PI_replicas}a. 
The overall prediction for the dilepton channel is shown in Fig. \ref{fig:DY_PI_replicas}b where we plot the central value and its theoretical error for the pure DY process (black line and band) and for the combined DY+PI process (red line and band). Clearly, we are in the presence of a very large systematic error that blows up quickly above $M_{ll}\ge 2.5$ TeV, as visible in Fig. \ref{fig:DY_PI_replicas}c. The relative PDF error is already well above 50\% at that mass scale and it reaches 500\% at $M_{ll}\ge 4$ TeV. At $M_{ll}\ge 4$ TeV, the central value of the PI process computed with NNPDF roughly equals the DY one. 

\begin{figure}[t]
\centering
\includegraphics[width=0.45\linewidth]{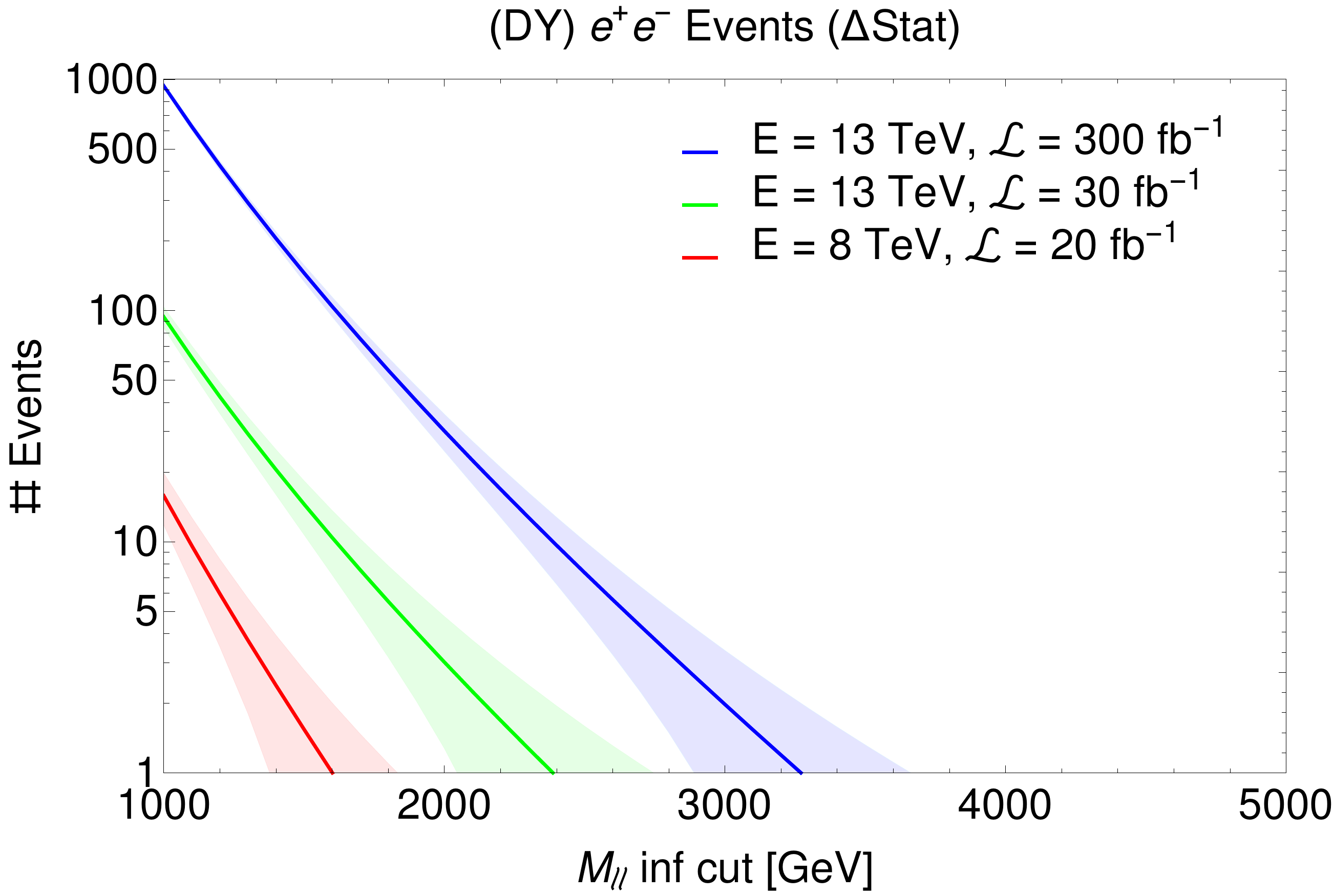}{(a)}
\includegraphics[width=0.45\linewidth]{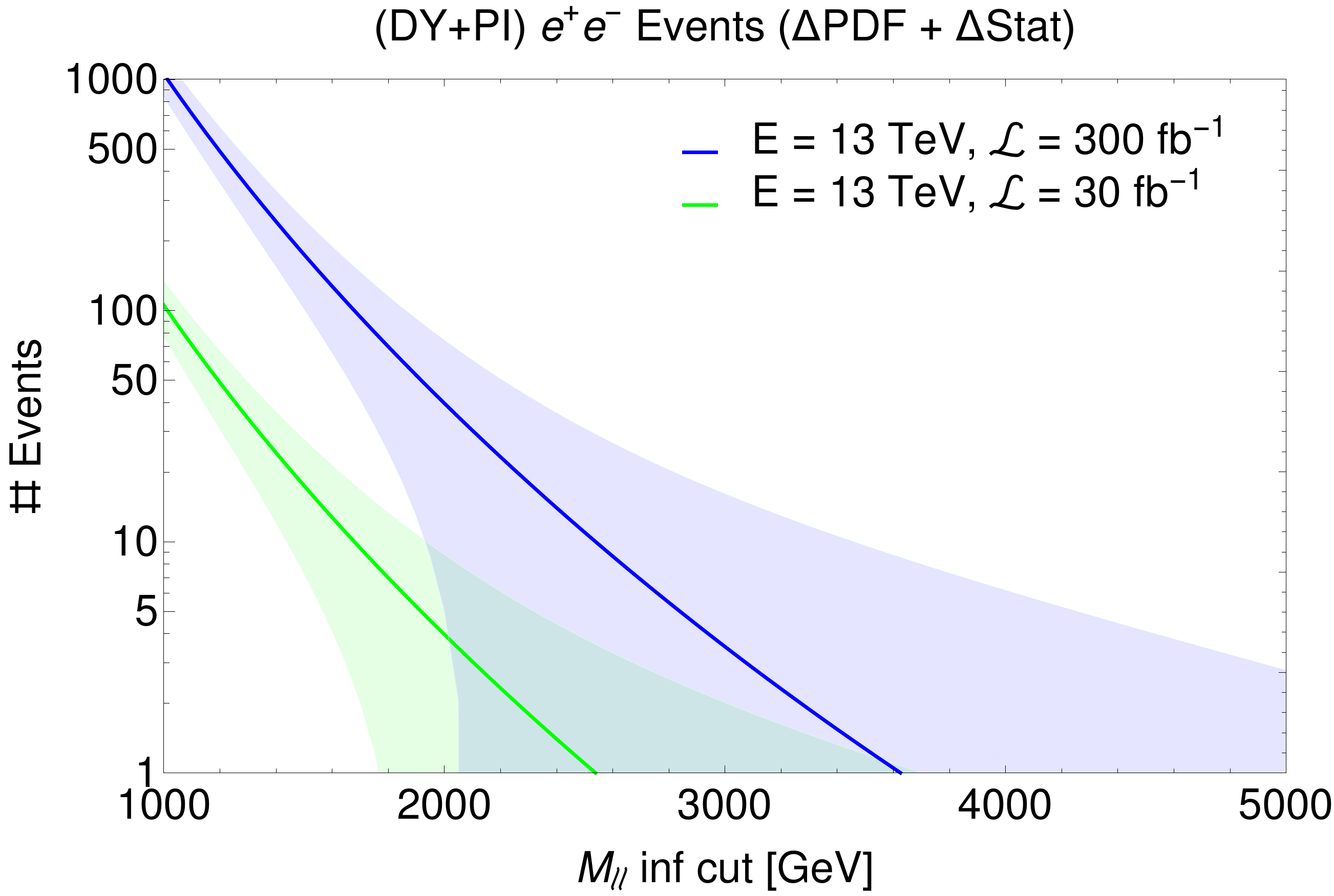}{(b)}
\caption{(a) Number of SM events expected in the dielectron channel from the DY process as a function of the lower cut on the dilepton invariant mass. 
The error bands include only the statistical error, as the PDF error is here sub-dominant.
(b) Same result with the inclusion of the PI contribution. 
In the error bands is now included the overall PDF uncertainty in addiction to the statistical error.
Standard acceptance cuts are applied ($|\eta_l| < 2.5$ and $p_T^l > 20~$GeV) as well as the declared efficiency of the electron channel \cite{Khachatryan:2014fba}. NNLO QCD corrections are accounted for in the DY term \cite{Hamberg:1990np}.}
\protect{\label{fig:Events}}
\end{figure}

\noindent
This means that the overall PI contribution, including the error, is a factor of four larger than the DY background considered by default in many experimental searches. The number of expected SM events and its uncertainty are shown in Fig. \ref{fig:Events} before (a) and after (b) the inclusion of the PI process and its uncertainty. We consider the past LHC RunI at 8 TeV and $L=20~fb^{-1}$ and the ongoing LHC RunII with two representative luminosities, $L=30~fb^{-1}$ and $L=300~fb^{-1}$. These plots confirm that the number of SM background events is very poorly determined. For the project luminosity $L=300~fb^{-1}$, the evaluation of the pure DY contribution could lead to the conclusion that the region above $M_{ll}=3.6$ TeV should be background free, while adding the PI contribution and its PDF error one would realize that the SM events could run over the spectrum up to $M_{ll}=5$ TeV and beyond.

\noindent
One can envisage two ways to keep under control this very large systematic uncertainty on the dilepton spectrum. The first one is resorting to a different observable, more robust against systematics. A possible candidate is AFB of the final state leptons \cite{Accomando:2015cfa,Accomando:2015pqa,Accomando:2015sun,Accomando:2015ava,Dittmar:1996my}.
For the default DY process, initiated by quarks (antiquarks) pairs, this observable has been shown to be mildly dependent on PDF uncertainties and robust against systematic errors, in general, being a ratio of cross sections. The second option would be implementing dedicated kinematical cuts in order to suppress the PI contribution, acting directly on the central value.
In the next two sections, we address these two possibilities.

\section{Forward-Backward Asymmetry}\label{sec:AFB}

It was already shown in Ref. \cite{Accomando:2015cfa} that systematic effects are strongly reduced in charge asymmetry observables like AFB. There, this was analyzed in the pure DY process and it was proven that the impact of the quark (antiquark) PDF error on the AFB determination is quite mild. In this paper, we show that the inclusion of the PI contribution does not alter this conclusion. The PI process itself does not produce (at tree level) any asymmetry in the angular distribution of the final state leptons, 
but it contributes to the total cross section that appears in the denominator of the AFB expression
\begin{equation}
A_{FB} = {{\sigma_F - \sigma_B}\over {\sigma_F + \sigma_B}},
\end{equation}
where $\sigma_F$ and $\sigma_B$ are the forward and the backward contributions to the total cross section, respectively. They are obtained by integrating the lepton angular distribution forward and backward with respect to the incoming quark direction. At a $pp$-collider like the LHC, the original quark direction is unknown. We thus reconstruct the AFB by extracting the quark direction from the kinematics of the dilepton system. Following the criteria of Ref. \cite{Dittmar:1996my}, we simulate the quark direction from the boost of the dilepton system with respect to the beam axis. The strategy is motivated by the fact that at the $pp$ LHC the dilepton events at high invariant mass come from the annihilation of either valence quarks with sea antiquarks or sea quarks with sea antiquarks. As the valence quarks carry away, on average, a much larger fraction of the proton momentum than the sea antiquarks, the boost direction of the dilepton system should give a good approximation of the quark direction. A leptonic AFB can thus be expected with respect to the boost direction. In contrast, the subleading number of dilepton events which originate from the annihilation of quark-antiquark pairs from the sea must be symmetric. As a measure of the boost, we define the dilepton rapidity
\begin{equation}
y_{ll} ={1\over 2} \ln \left ({{E + P_z}\over{E - P_z}}\right ),
\end{equation}
where $E$ and $P_z$ are the energy and the longitudinal momentum of the dilepton system, respectively. We identify the quark direction through the sign of $y_{ll}$. In this way, one can define the reconstructed forward-backward asymmetry, from now on called $AFB^*$.

\begin{figure}[t]
\centering
\includegraphics[width=0.45\linewidth]{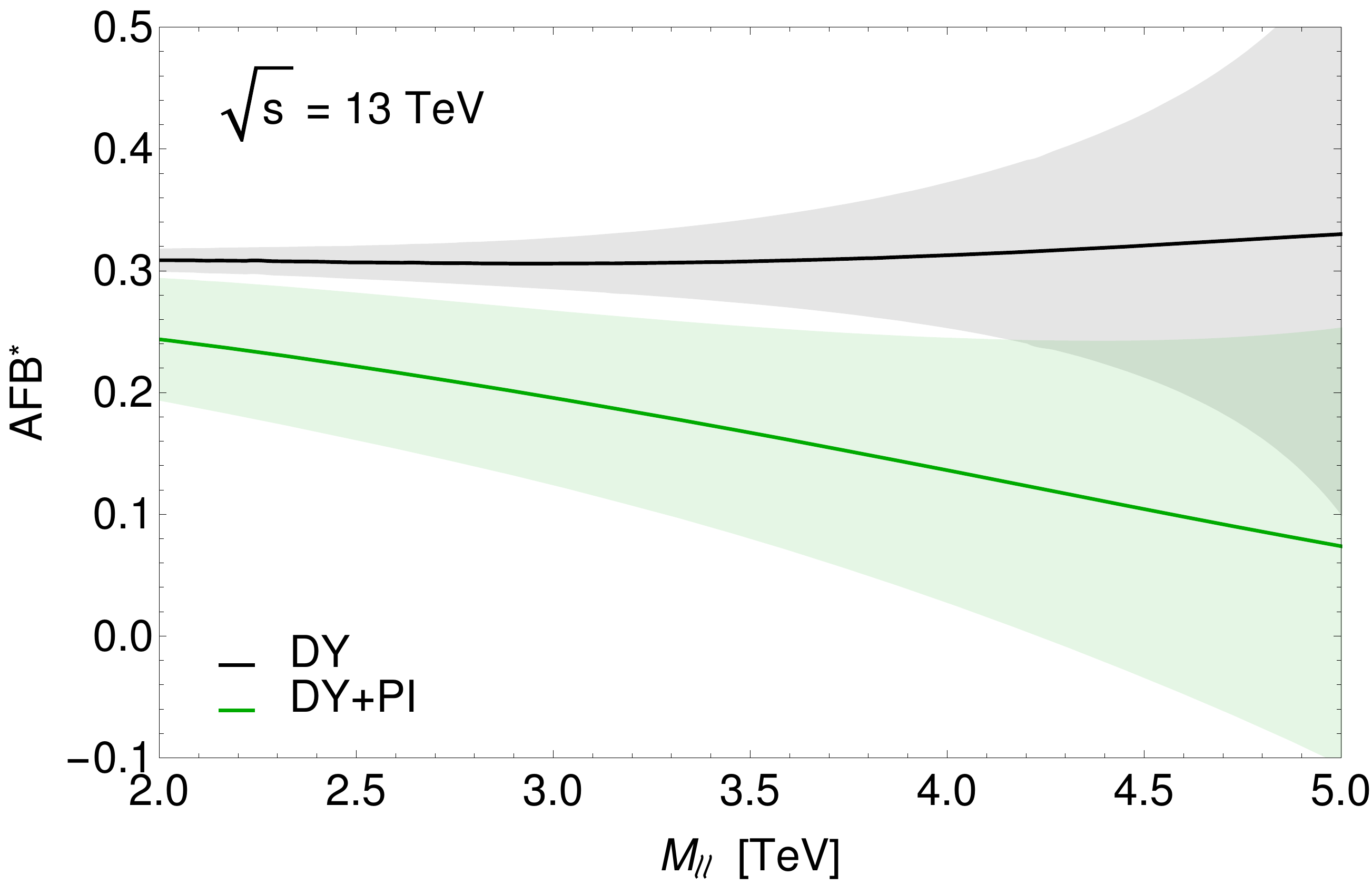}
\includegraphics[width=0.45\linewidth]{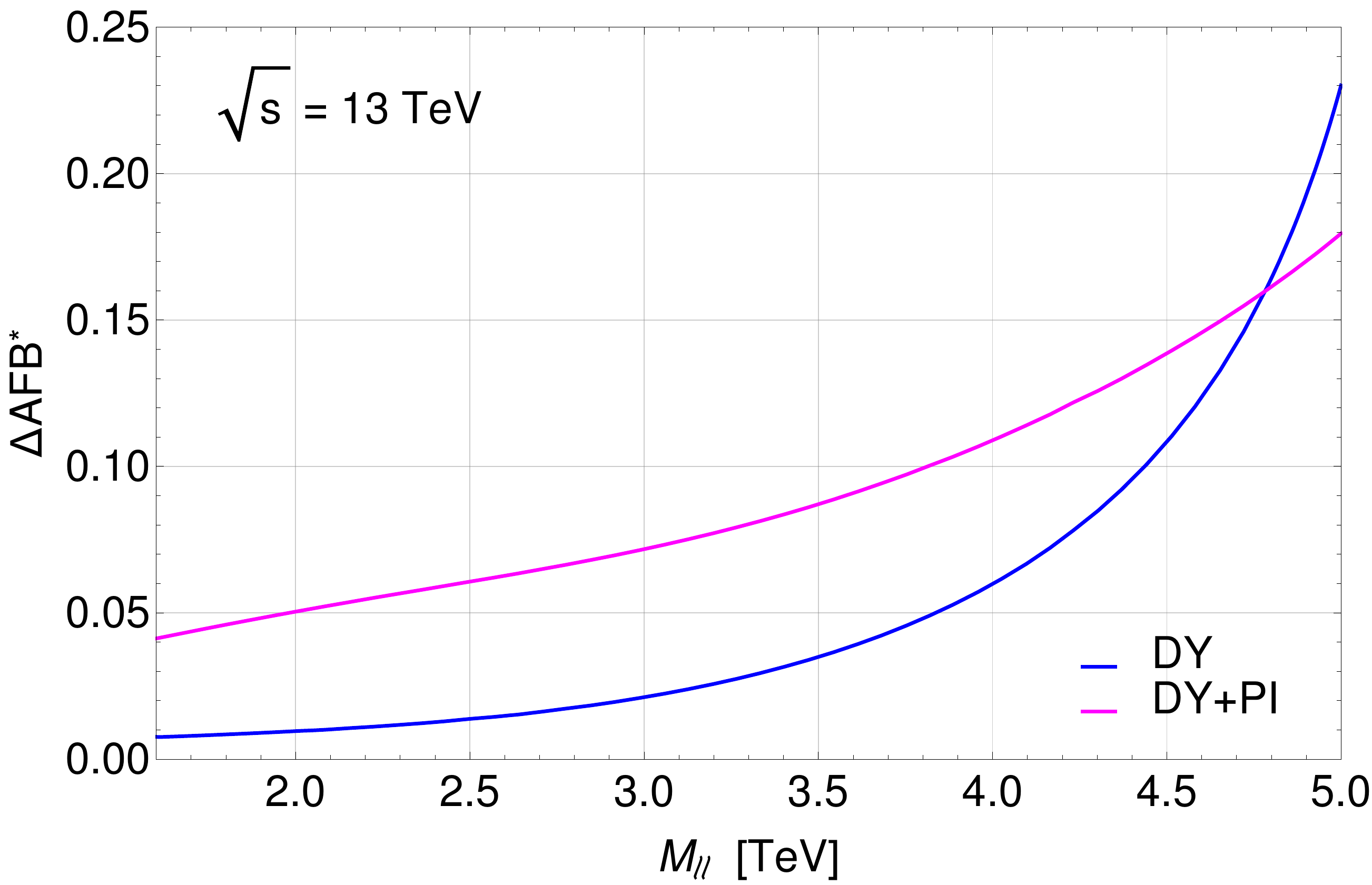}
\caption{(a) Reconstructed AFB for the DY process and the combined (DY + PI) process. 
(b) Absolute PDF error on the reconstructed AFB observable with and without the inclusion of the PI process. Standard acceptance cuts are applied ($|\eta_l| < 2.5$ and $p_T^l > 20~$GeV).}
\protect{\label{fig:AFB_error}}
\end{figure}

\noindent
In Fig. \ref{fig:AFB_error}a, we plot the reconstructed AFB within the SM with and without the PI contribution and its error. The central value of the $AFB^*$ decreases when we include the PI process, as expected, while the $AFB^*$ uncertainty increases only mildly. The stability of the $AFB^*$ shape against PDF uncertainties is even more visible in Fig. \ref{fig:AFB_error}b where we compare the $AFB^*$ PDF uncertainty with and without the PI contribution. The reconstructed AFB could then be used as a valuable supplementary observable to help understand  the experimental results coming from the analysis of the dilepton spectrum.

\section{On the effect of kinematical cuts}\label{sec:cuts}

In the DY channel, the lepton pairs are generated by the exchange of the SM $\gamma$ and $Z$-boson produced in $s$-channel. The PI process instead involves $t,u$-channel exchanges of a light lepton. The PI matrix element is affected by a collinear divergence that can be regulated by introducing a cut-off. This problem is naturally solved once we consider the detector acceptance. 
Imposing the fiducial cuts of a typical LHC detector, $|\eta_l| < 2.5$ and $p_T^l > 20~$GeV, the collinear divergence is controlled adequately.

\begin{figure}[t]
\centering
\includegraphics[width=0.45\linewidth]{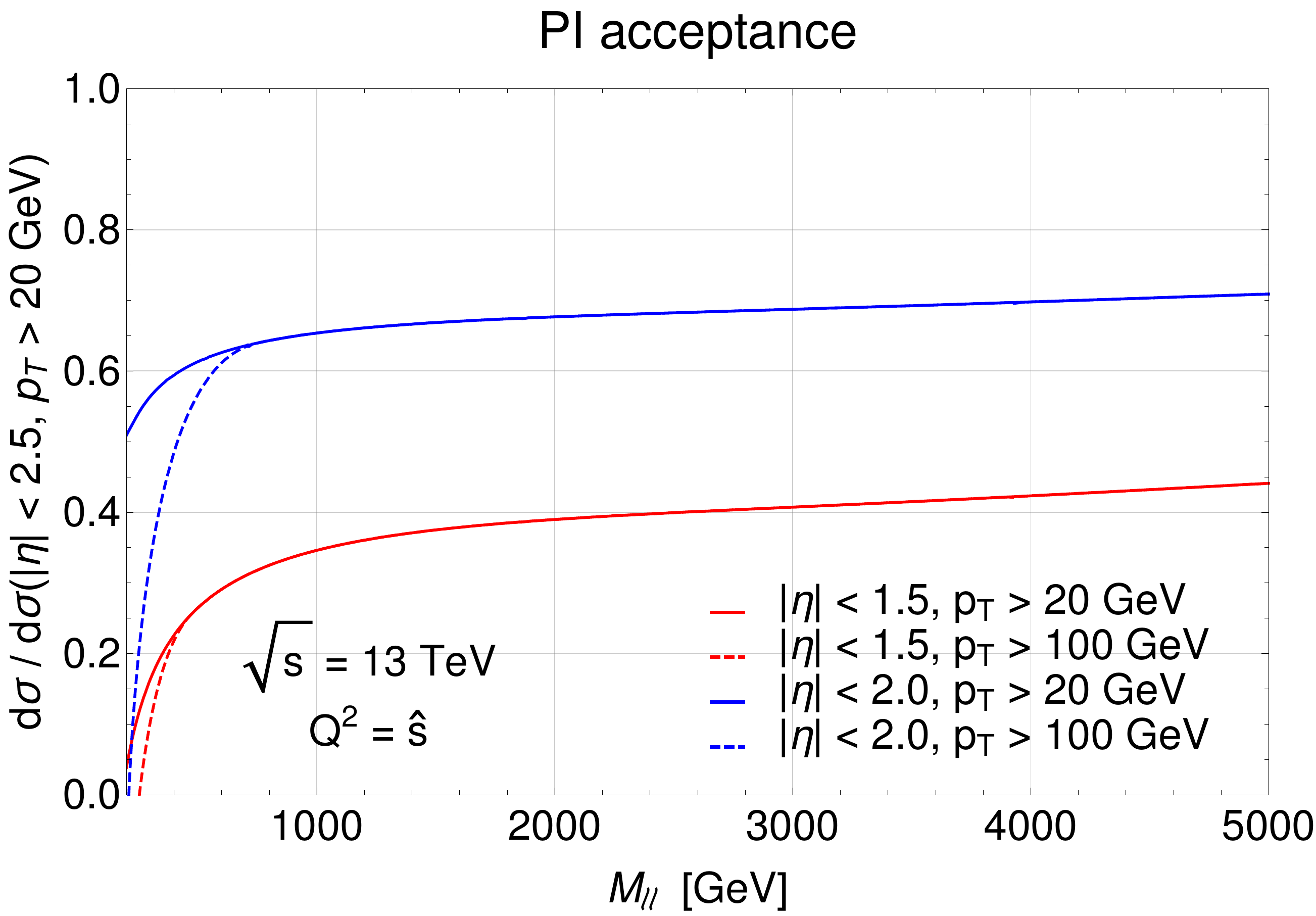}{(a)}
\includegraphics[width=0.45\linewidth]{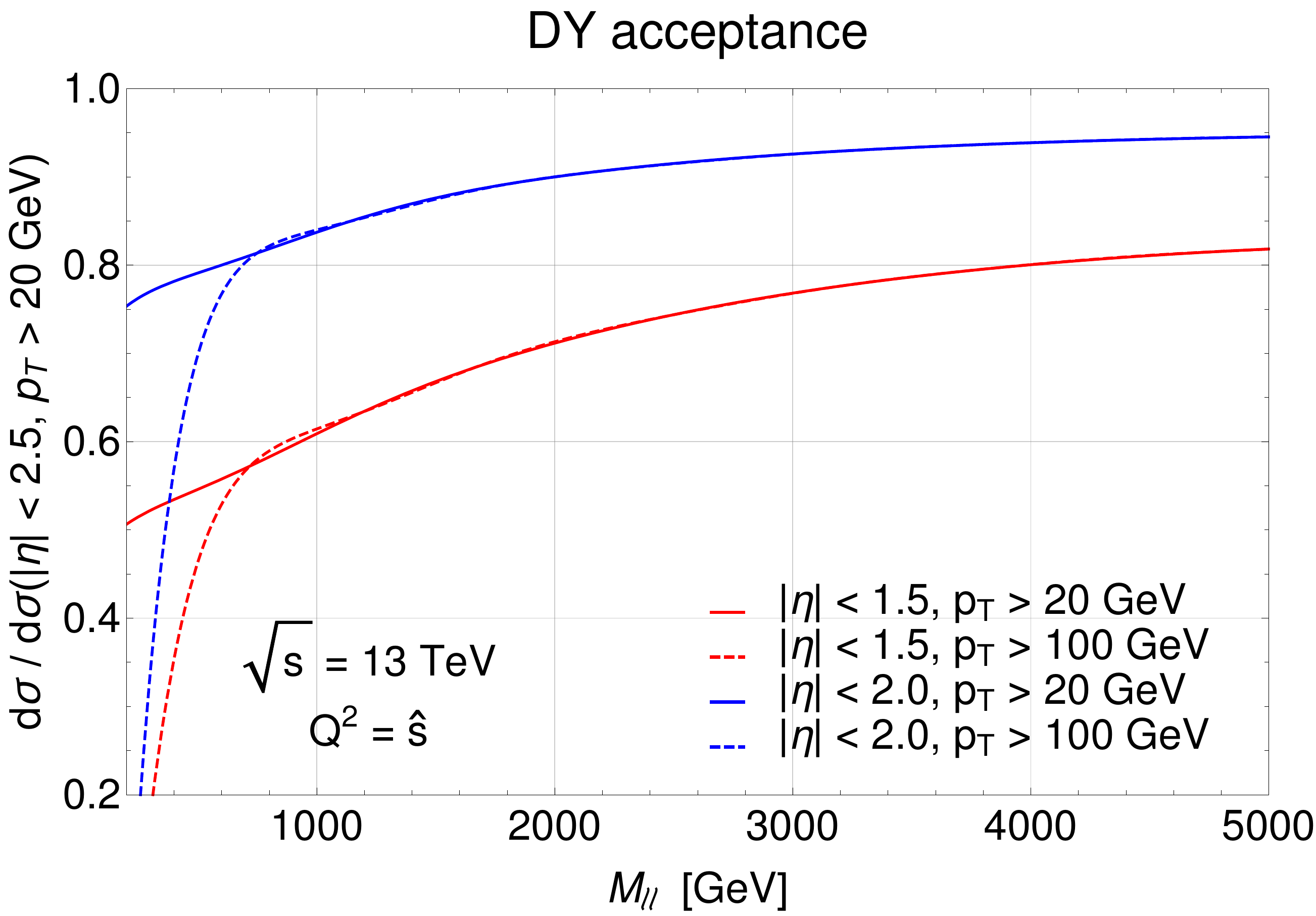}{(b)}
\caption{(a) The acceptance of the PI contribution to the dilepton spectrum for different cuts in $\eta_l$ and $p_T^l$ with respect to the differential cross section obtained with the standard acceptance cuts: $|\eta_l| < 2.5$ and $p_T^l > 20~$GeV. 
(b) Same for the DY process.
}
\protect{\label{fig:eta_pT}}
\end{figure}

\noindent
In the attempt to suppress the expected PI background, which is affected by very large  uncertainties and can pollute BSM searches at high invariant masses, we analyse the behaviour of the acceptance of the PI process versus the DY one with respect to different kinematical cuts that could be experimentally applied. We consider different angular and transverse momentum cuts.
In particular, the effect of the $\eta_l$ and $p_T^l$ cuts are shown in Figs. \ref{fig:eta_pT}a and \ref{fig:eta_pT}b for the PI and DY contributions, respectively. The first one appears more sensitive to the pure angular cuts encoded in the $\eta_l$ constraint. For $M_{ll} > 2~$ TeV, the cut on the lepton transverse momentum is practically ineffective. Imposing an angular cut of $|\eta_l| < 1.5$ can reduce the PI effect by about 60\%. The same cut decreases also the DY differential cross section by roughly 25\% in the region of interest, say, $M_{ll} > 2.5~$ TeV. 
So, despite the increasing virtuality of the fermion exchanged in $t,u$-channels, the PI process does not get suppressed much by the $\eta_l$ cut with respect to the DY process. 
 
\noindent
In essence, the number of PI events can be reduced imposing stronger angular cuts, but the benefits of this might not justify the consequent loss of statistics in the DY channel as this could contain BSM physics. An alternative strategy to adopt would be to include the PI and its theoretical error in the SM background estimate.

\section{$Z^\prime$-boson searches and the SM background}\label{sec:BSM}

The scenario we have pictured does not look like a friendly environment for the precise detection of new physics signals. 
The theoretical uncertainty on the SM prediction is dominated by systematic effects that will be mostly influential with increasing the integrated luminosity.
Searches for new heavy neutral resonances are better performed in the dilepton channel, but the theoretical inputs intervening in the experimental procedures should be ameliorated in the light of the results we have shown. 
BSM searches are indeed performed simulating the SM background via a functional form whose parameters are fitted to the MC predictions. 
The net result is then normalized to the data in an invariant mass region, appropriately chosen, on the left-hand side of the hypothetical $Z^\prime$-boson mass. 
The shape or slope of the DY dilepton spectrum can be modified significantly by the PI contribution, including its error. 
The combined (DY + PI) background decreases much less steeply with the invariant mass than the default DY. 
So does the overall number of expected SM events. 
For invariant masses above $M_{ll}=$ 3.5 TeV, QED effects might increase the number of events from one (as expected in pure DY production) to twenty (including the uncertainties). 
A reshaping of the SM background is thus advisable in order to perform an accurate likelihood fitting procedure.

\noindent
In this context, both the resonant search for a new heavy neutral gauge boson and the non-resonant searches for contact interactions, to cite an example, can be strongly affected by the PI contribution. 
In the resonant case, the incorrect estimate of the SM background events would lead to either a pre-discovery enthusiasm while in presence of a simple fluctuation or to mis-estimate the $Z^\prime$-boson mass bounds. 
In non-resonant searches we would encounter more difficulties. The preferred experimental strategy to deal with such scenarios is the counting strategy. 
This consists in imposing a lower cut on the dilepton invariant mass and in integrating from there over the whole dilepton spectrum, searching for an excess of events. 
Such an excess could appear as a shoulder over the expected SM background in case of wide resonances or as a plateau standing over the SM background in case of contact like BSM interactions. 
As we have shown in the previous section, this prescription is actually dominated by systematic uncertainties coming from the photon PDFs. 
Again an excess of events can be here mis-interpreted as new physics while in presence of purely QED effects.

\noindent
In Sect. \ref{sec:AFB}, we have shown that the forward-backward charge asymmetry is much less affected by PDFs uncertainties than the dilepton spectrum within the SM. 
Following this path, we want now to extend our analysis and see how new physics signals, which might appear in these two observables, would respond to the noise coming from the PI contribution and its uncertainty at the ongoing LHC RunII. 
We focus on the $Z^\prime$-boson physics and select as benchmarks the $E_6^\chi$ model \cite{Langacker:2008yv,Erler:2009jh,Nath:2010zj,Accomando:2010fz}, representative of narrow width $Z^\prime$ bosons, and the GSM-SSM \cite{Altarelli:1989ff} that is representative of wide resonances. 

\subsection{Narrow width $Z^\prime$-boson}

\begin{figure}[t]
\centering
\includegraphics[width=0.45\linewidth]{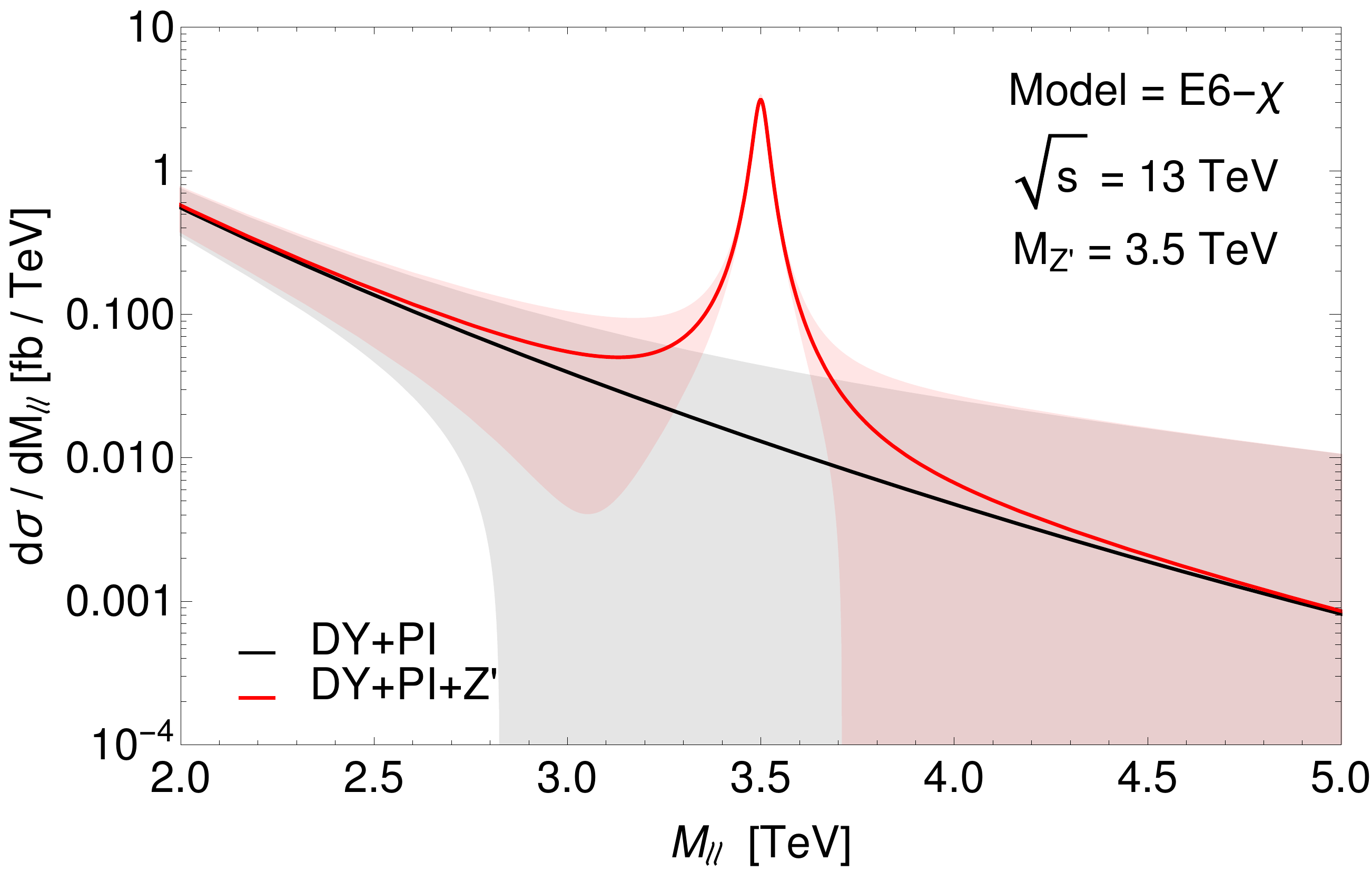}{(a)}
\includegraphics[width=0.45\linewidth]{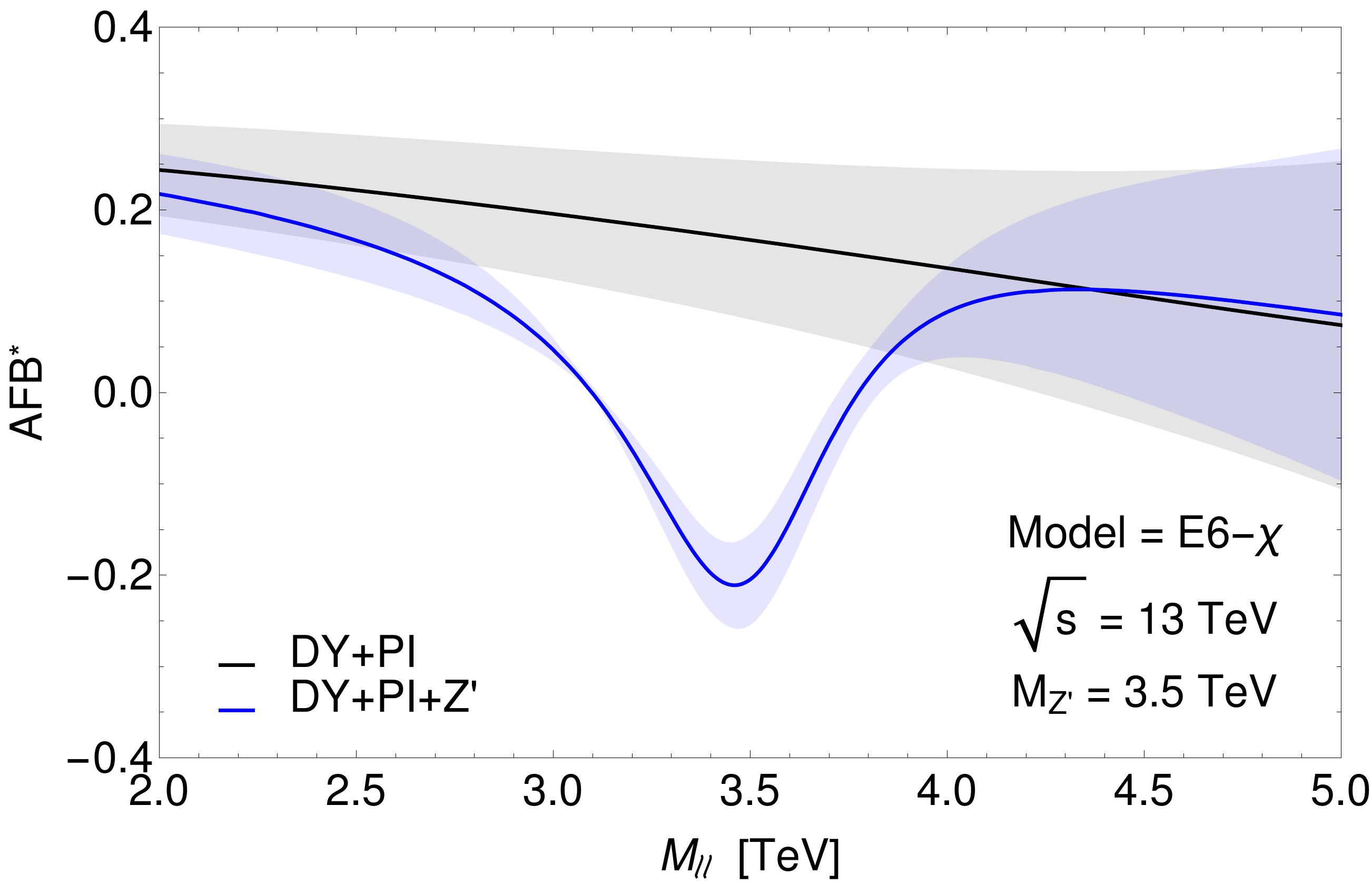}{(b)}
\caption{(a) Differential cross section predicted within the $E_6^\chi$ model with a $Z^\prime$-boson of mass $M_{Z'}=$ 3.5 TeV. 
The black solid line represents the combined (DY + PI) background including PDFs uncertainty and the red line the full contribution including the $Z^\prime$ signal.
(b) Reconstructed AFB distribution for the same benchmark including PDF uncertainty. 
The line color code is the same as in (a). Standard acceptance cuts are applied ($|\eta_l| < 2.5$ and $p_T^l > 20~$GeV).}
\protect{\label{fig:E6_chi}}
\end{figure}

In this section, we consider the extra heavy $Z^\prime$-boson predicted by the $E_6^\chi$ model \cite{Langacker:2008yv,Erler:2009jh,Nath:2010zj,Accomando:2010fz}, which is characterized by a narrow width. 
The present mass bound for this particle is $M_{Z'}\ge$ 2700 GeV \cite{Accomando:2015cfa}. 
We compare the dilepton spectrum with the reconstructed AFB for this scenario, with particular emphasis on the theoretical error due to QED effects. 
The impact of the inclusion of the PI process in the SM background is shown in Fig. \ref{fig:E6_chi} where we display the differential cross section in the dilepton invariant mass (a) and the reconstructed AFB (b) in the same variable for $M_{Z'}=$ 3.5 TeV. 
The error bands in the plots represent the PDF uncertainties on the corresponding observables.

\begin{figure}[t]
\centering
\includegraphics[width=0.45\linewidth]{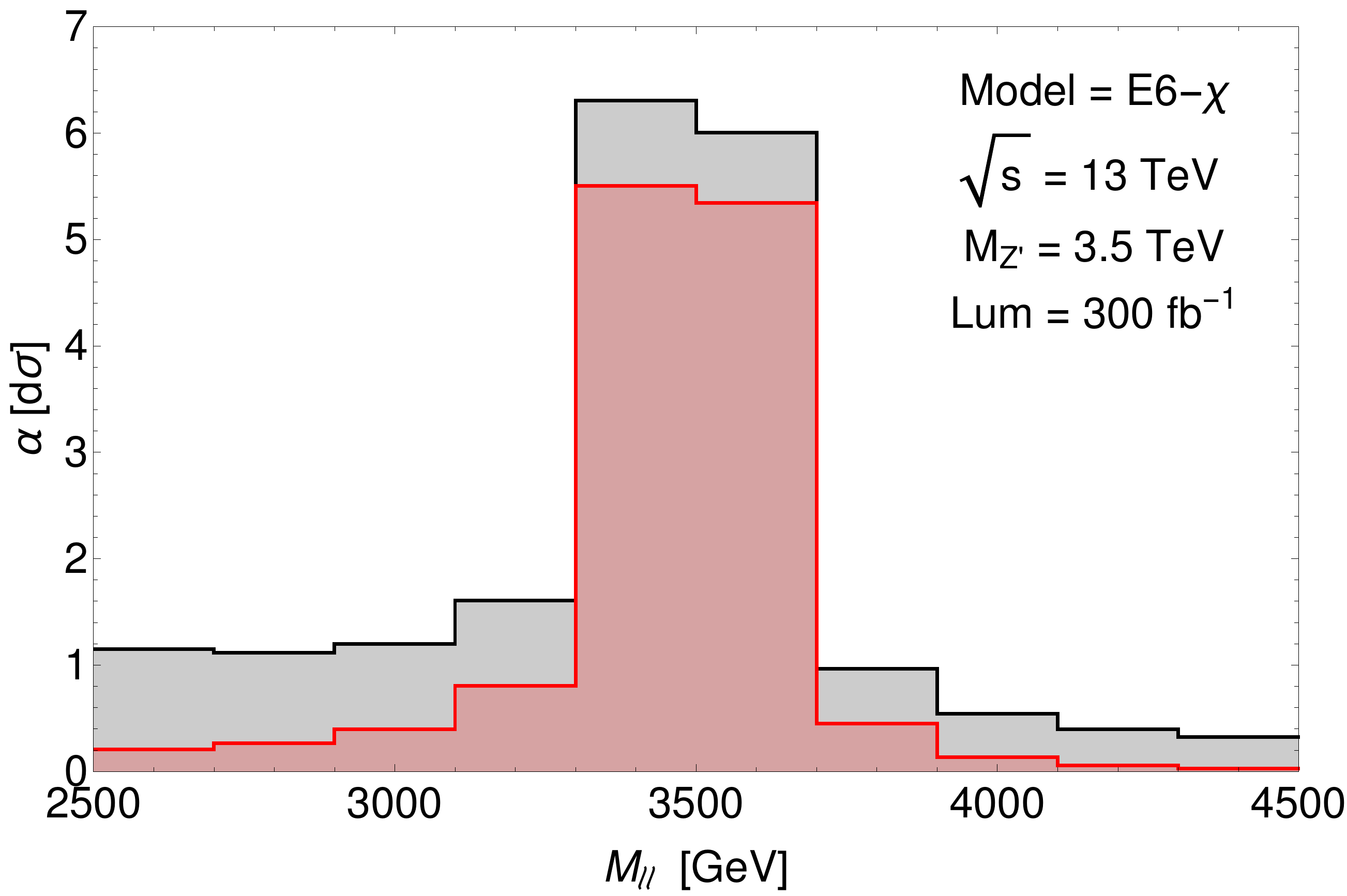}{(a)}
\includegraphics[width=0.45\linewidth]{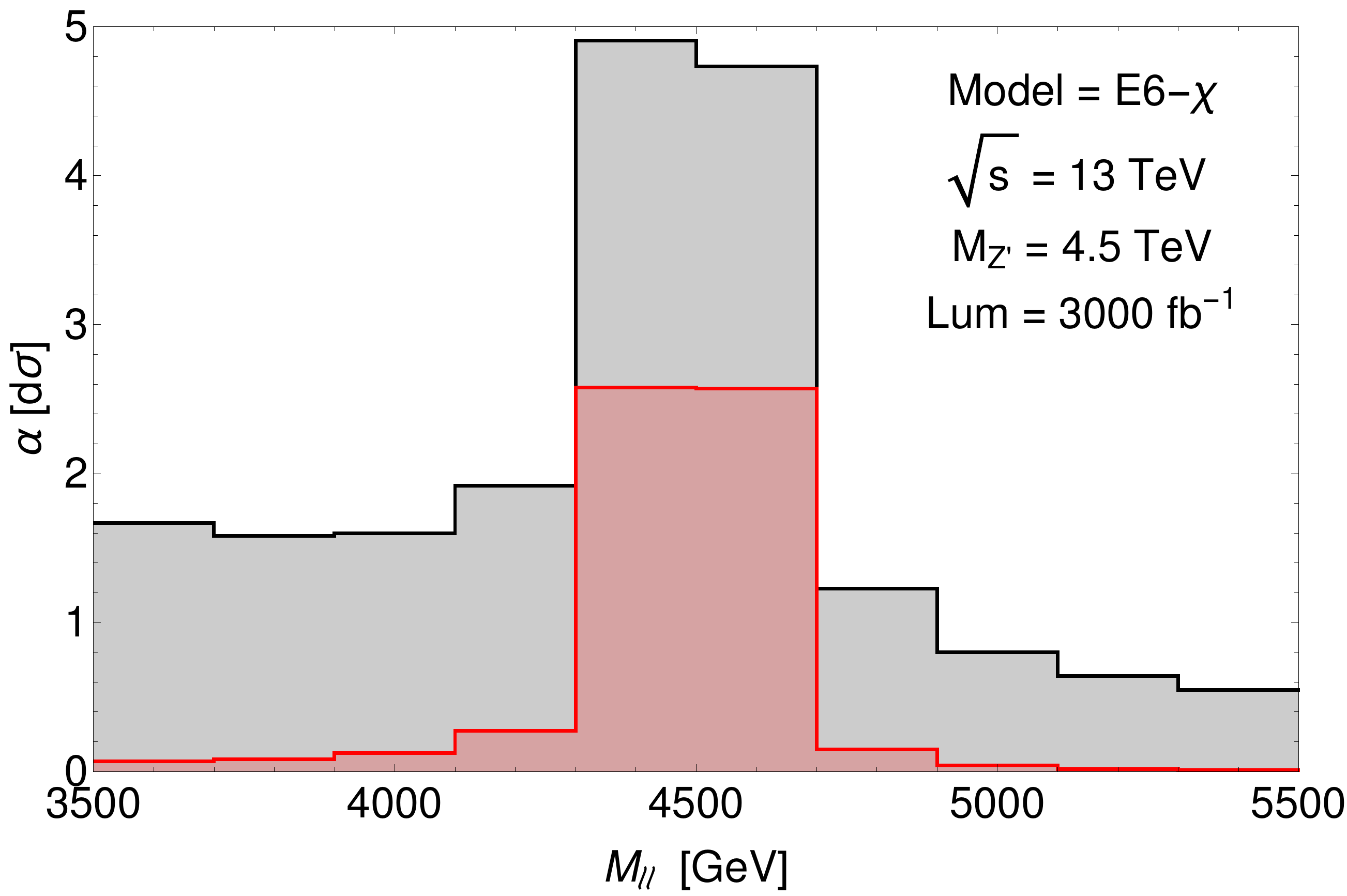}{(b)}
\includegraphics[width=0.45\linewidth]{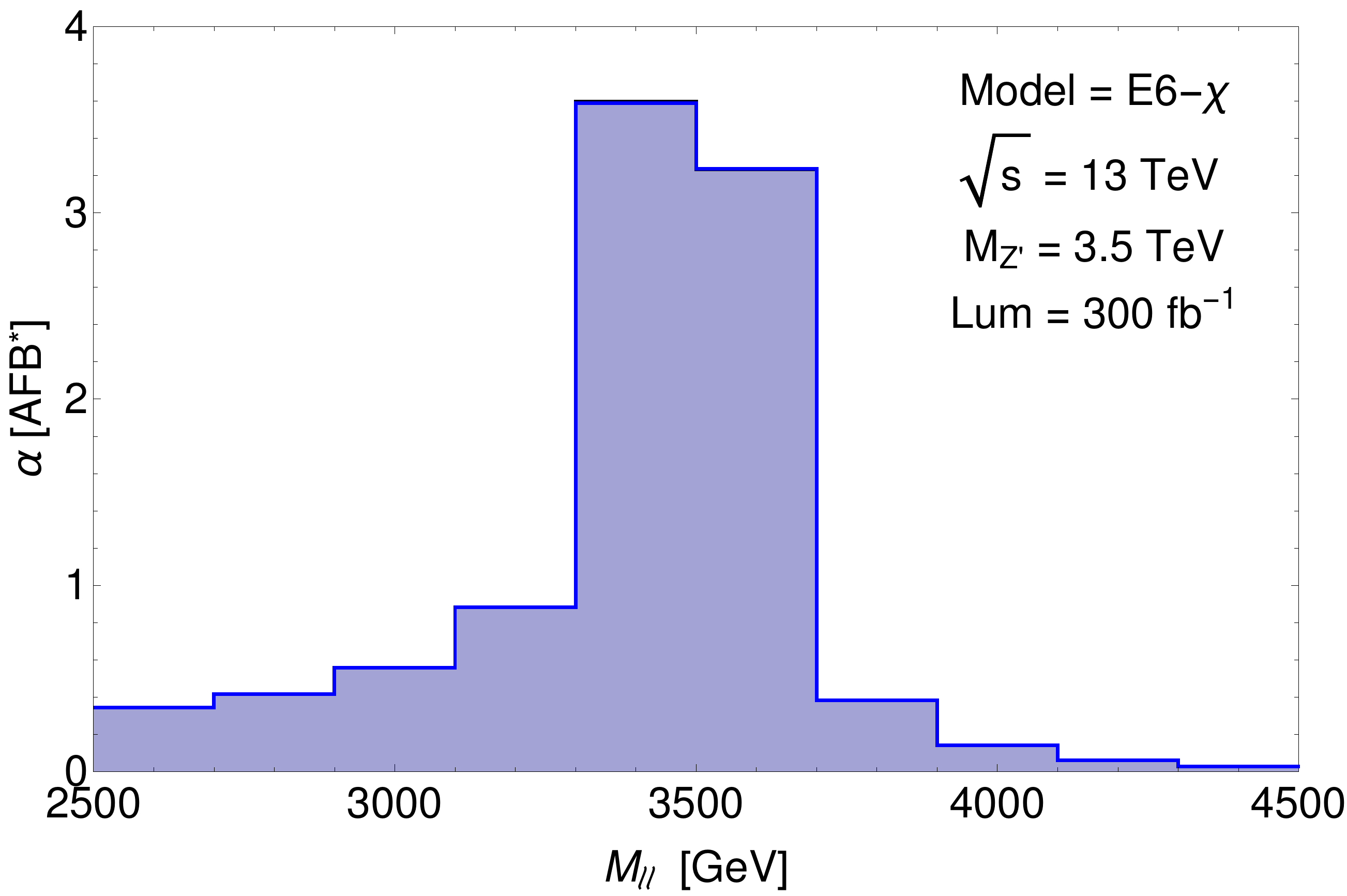}{(c)}
\includegraphics[width=0.45\linewidth]{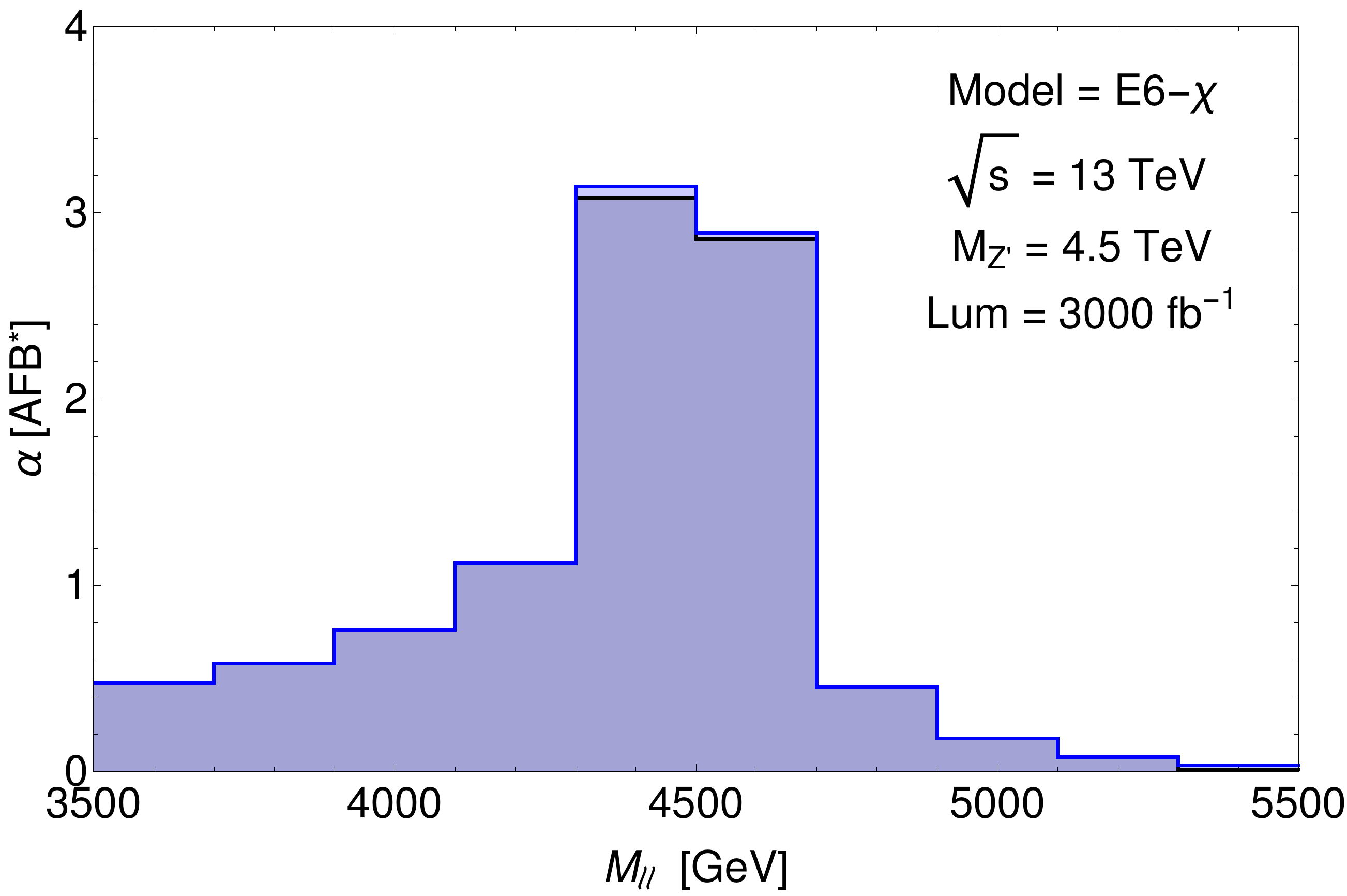}{(d)}
\caption{(a) Differential cross section significance within the $E_6^\chi$ model with a $Z^\prime$ of mass $M_{Z'}=$ 3.5 TeV. 
The black line represents the case where only the default DY process is accounted for as a SM background. 
The red line represents the case where the combined (DY + PI) process is taken into account as SM background. 
The integrated luminosity is $L=300~fb^{-1}$. 
(b) Same as (a) for a $Z^\prime$ with $M_{Z'}=$ 4.5 TeV and an integrated luminosity $L=3~ab^{-1}$.
(c) Significance of the reconstructed AFB distribution with and without the PI contribution, for a $Z^\prime$ boson of mass $M_{Z'}=$ 3.5 TeV and an integrated luminosity $L=300~fb^{-1}$. 
(d) Same as (c) for a $Z^\prime$ with $M_{Z'}=$ 4.5 TeV and an integrated luminosity $L=3~ab^{-1}$. Standard acceptance cuts are applied ($|\eta_l| < 2.5$ and $p_T^l > 20~$GeV) as well as the declared efficiencies of the electron and muon channels \cite{Khachatryan:2014fba}. NNLO QCD corrections are accounted for in the DY term \cite{Hamberg:1990np}. The overall significance is the combination of the significances in the two lepton channels. The binning has been chosen to represent an average of the two channel resolutions.}
\protect{\label{fig:E6_chi_sig}}
\end{figure}

\noindent
The sizeable uncertainty generated by the inclusion of the QED effects is evident from the plots. 
We can moreover conclude that the reconstructed AFB is more robust against PDF errors than the spectrum, also in the instance of new physics. 
The inclusion of the photon induced lepton pairs  and their PDF uncertainties is crucial in the estimate of the significance of the BSM signal. 
In Fig. \ref{fig:E6_chi_sig} we consider the two cases where we (correctly) include the PI contribution in the SM background, quoting its uncertainties in the overall error, (colored lines) and where the PI events are considered as part of the new physics signal and the QED PDFs uncertainty in not included in the overall error (black lines). 
This comparison is shown for the the dilepton spectrum (upper plots) and the reconstructed AFB (lower plots), for two different values of the luminosity $L=300~fb^{-1}$ (the project luminosity that will be reached in a three years time) and $L=3~ab^{-1}$ (the project value of the high luminosity LHC upgrade). 
The significance is defined as

\begin{equation}
\alpha = \frac{|S-B|}{\Delta(S+B)}
\end{equation}
where $S$ represents the BSM signal and $B$ is the expected SM background. 
The overall uncertainty is the quadratic sum of the statistical and PDF errors, $\Delta(S+B)^2 = \Delta_{\rm stat}^2 + \Delta_{\rm PDF}^2$. 
The PDF error has been estimated as described in Sect. \ref{sec:PDFerror}, while the statistical error for the two observables is calculated as

\begin{align} 
&\Delta_{\rm stat}^{d\sigma} = \sqrt{N}\\ 
&\Delta_{\rm stat}^{\rm AFB^*} = \sqrt{\frac{1-AFB^{*2}}{N}}
\end{align}
where $AFB^*$ is the reconstructed AFB and $N$ is the total number of expected SM events. 
Even if quite basic, this estimate of the total error and consequently of the significance gives already a fair perspective of the impact of the PI contribution on the interpretation of BSM searches. 

\noindent
As one can see from the upper plots in Fig. \ref{fig:E6_chi_sig}, with increasing the luminosity the possibility of detecting a new $Z^\prime$-boson at higher mass in the dilepton spectrum would in principle grow, owing to the reduced statistical error. 
However, the PI contribution and its theoretical error, both increasing with the energy scale, cap this potential enhancement. 
In this respect, the reconstructed AFB has the ability to cope with the QED theoretical error much better. 
The PI contribution has no appreciable effect on the significance, as shown in the two lower plots in Fig. \ref{fig:E6_chi_sig}. 
Similar results will be found in next section, where we discuss the broad resonance case. 

\subsection{Wide $Z^\prime$-boson}

Experimental searches for non-resonant objects in the invariant mass distribution are usually performed adopting a counting strategy approach. 
That means imposing a lower cut on the dilepton invariant mass and summing over all events from there on. 
One thus compares the observed number of events with the theoretical expectation. 
For any meaningful interpretation of BSM searches, it is then of great importance to have a precise determination of the SM background in magnitude and shape. 
In this case, the photon induced dilepton events play a major role. 
They indeed become relevant for $M_{ll}\ge$ 3 TeV, especially if one considers their PDF uncertainties.

\noindent
In this section, we consider a broad $Z^\prime$-boson with couplings as predicted by the  GSM-SSM  \cite{Altarelli:1989ff}. 
We take the ratio $\Gamma / M_{Z^\prime} = 20\%$. 
Our choice is dictated by the fact that this model is often used as benchmark by the experimental collaborations. 

\begin{figure}[t]
\centering
\includegraphics[width=0.45\linewidth]{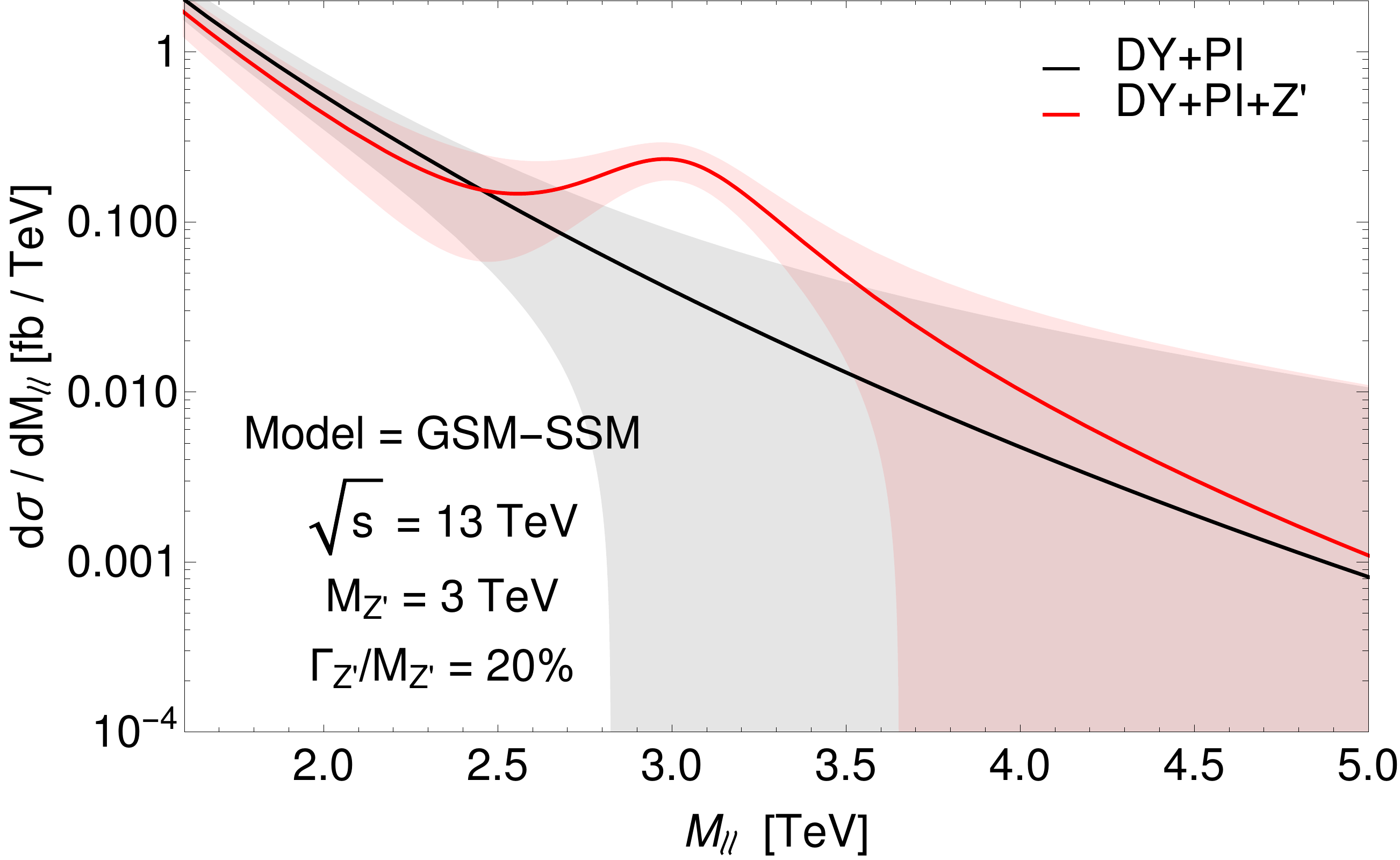}{(a)}
\includegraphics[width=0.45\linewidth]{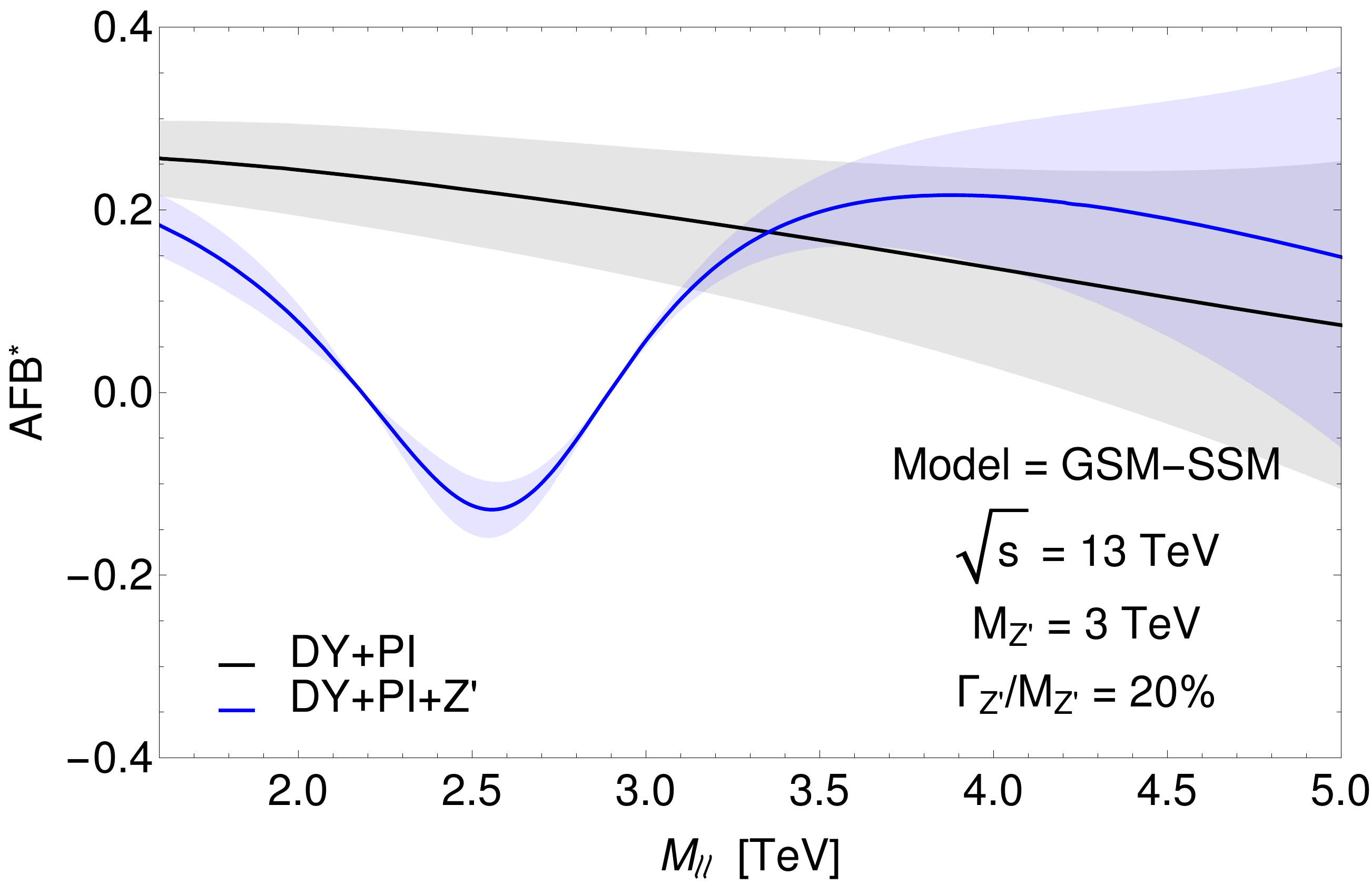}{(b)}
\caption{(a) Differential cross section predicted within the  GSM-SSM  model with a $Z^\prime$-boson of mass $M_{Z'}=$ 3 TeV and $\Gamma / M_{Z^\prime} = 20\%$. 
The black dashed line represents the pure DY background as default reference, the black solid line the combined (DY + PI) background including the PDFs uncertainty and the red line the full contribution including the $Z^\prime$ signal.
(b) Reconstructed AFB distribution for the same benchmark including PDF uncertainty. 
The line color code is the same as in (a). Standard acceptance cuts are applied ($|\eta_l| < 2.5$ and $p_T^l > 20~$GeV).}
\protect{\label{fig:SSM_bench}}
\end{figure}

\noindent
Differential cross section and reconstructed AFB distribution in the dilepton invariant mass within this model are shown in Fig. \ref{fig:SSM_bench}, where the error bands represent here the PDF uncertainties. 
In contrast to the previous narrow-width case, now the noise on the spectrum coming from the SM dilepton production with the inclusion of QED effects is much larger. 
Compared to the default DY background, represented by the dashed black line in Fig. \ref{fig:SSM_bench}a, the full SM background coming from the combined (DY+ PI) process and depicted by the gray region around the central value (black solid line) has a sizeably different magnitude and shape. 
Such shape could easily fake either a broad resonance or a non-resonant type of new physics signal (like the well studied contact interactions). 
Once again, the reconstructed AFB looks much more solid in presence of QED effects and related theoretical uncertainties as shown in Fig. \ref{fig:SSM_bench}b. 
This appears also in the estimate of the significance. 

\begin{figure}[t]
\centering
\includegraphics[width=0.45\linewidth]{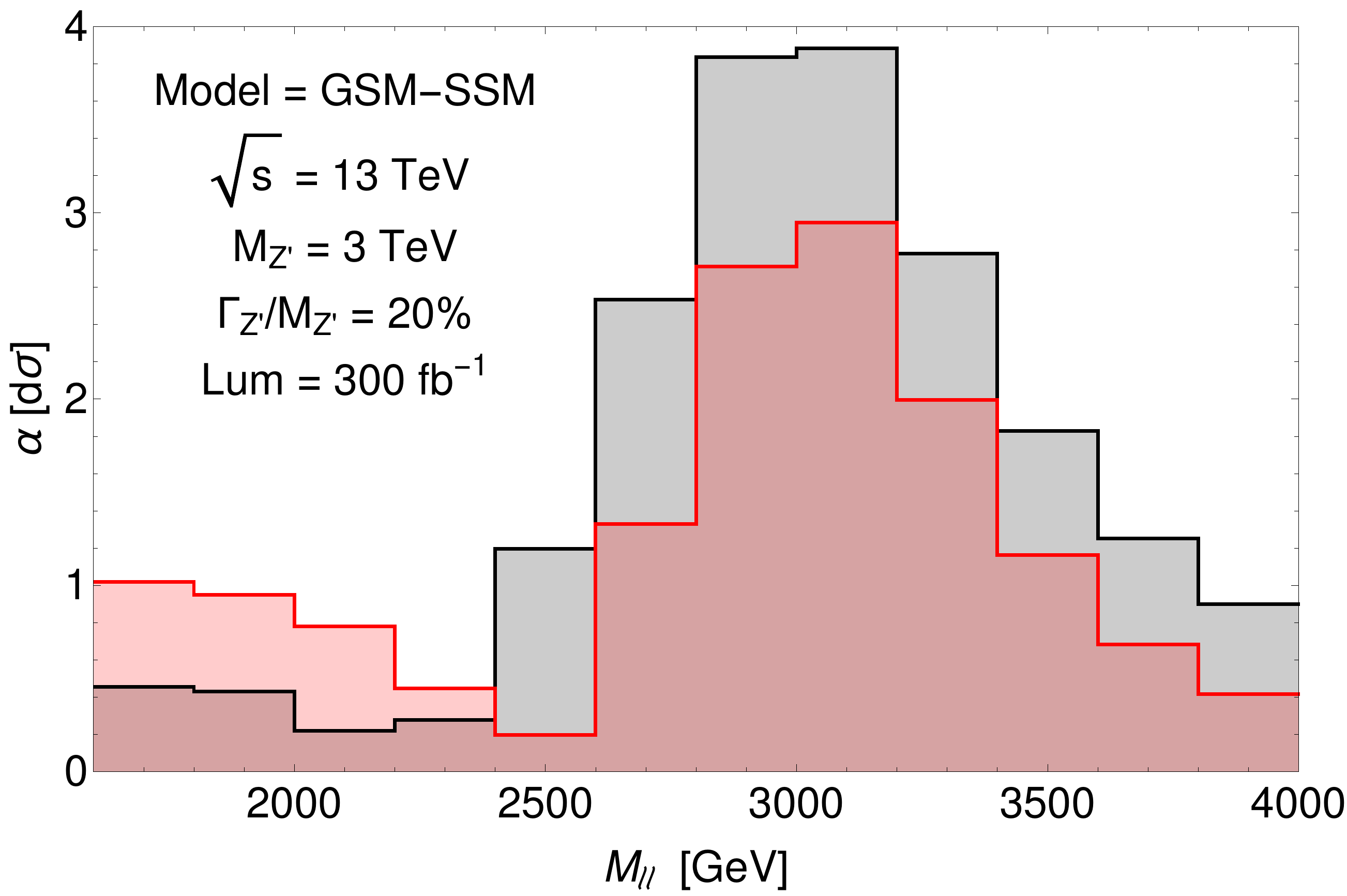}{(a)}
\includegraphics[width=0.45\linewidth]{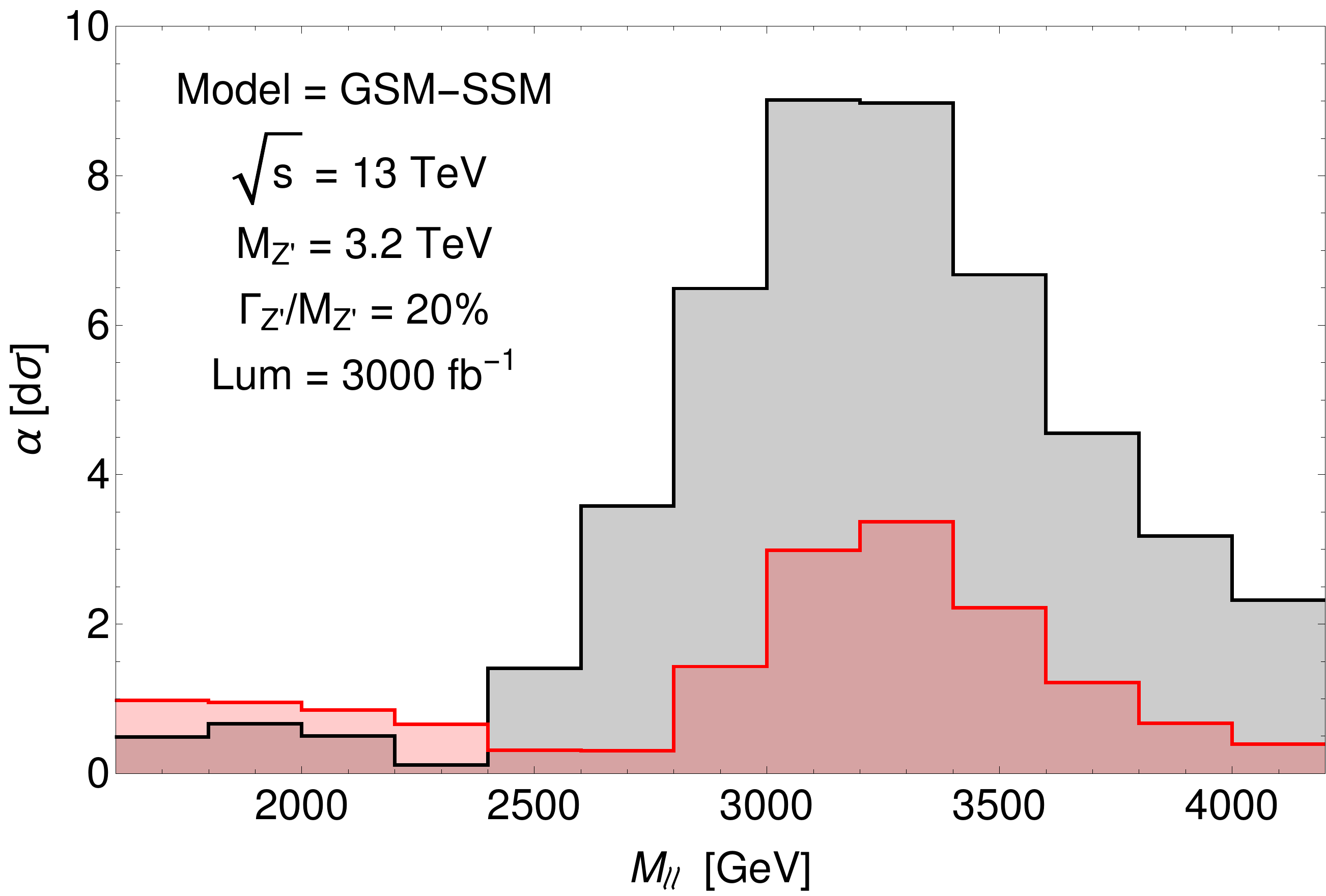}{(b)}
\includegraphics[width=0.45\linewidth]{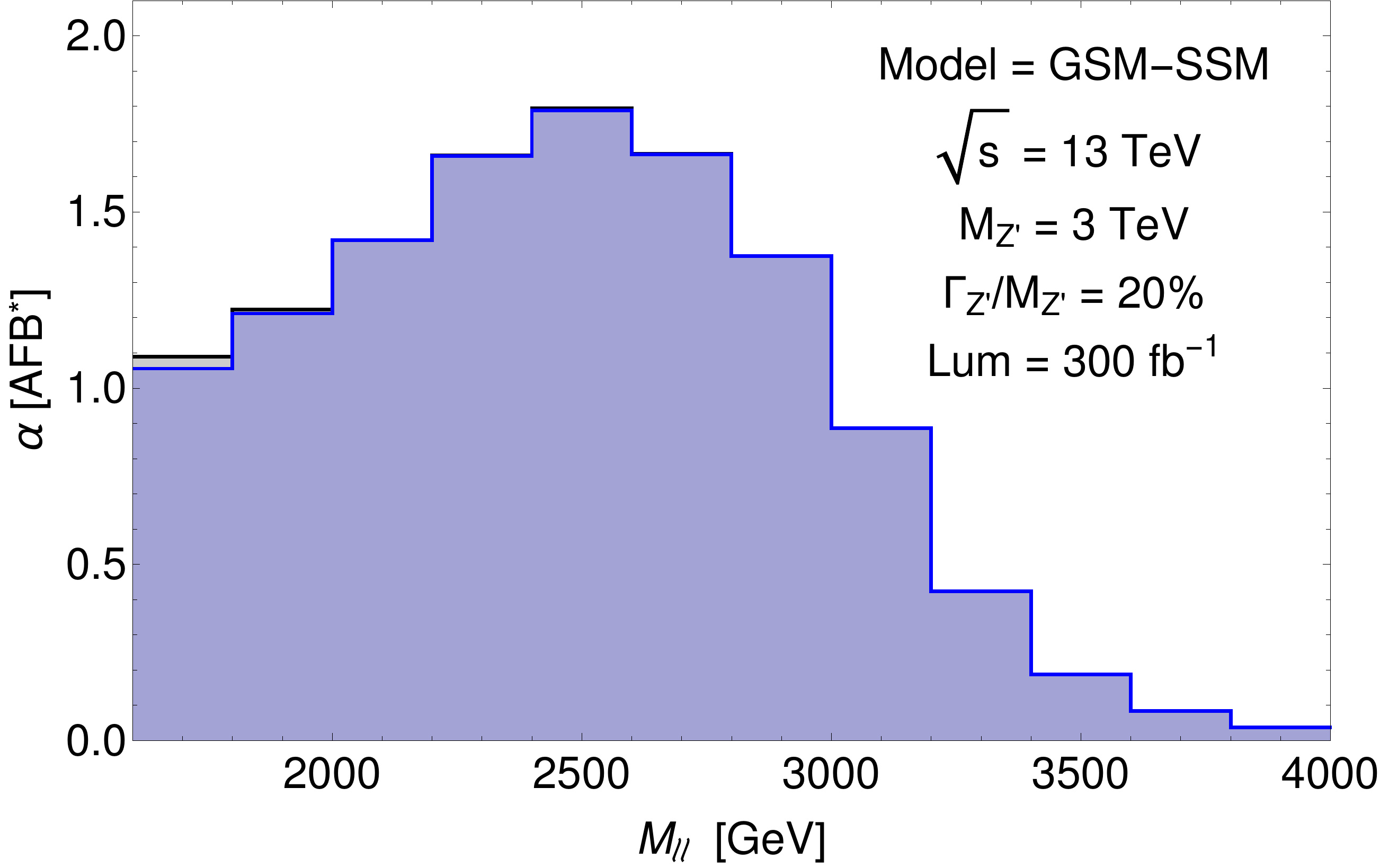}{(c)}
\includegraphics[width=0.45\linewidth]{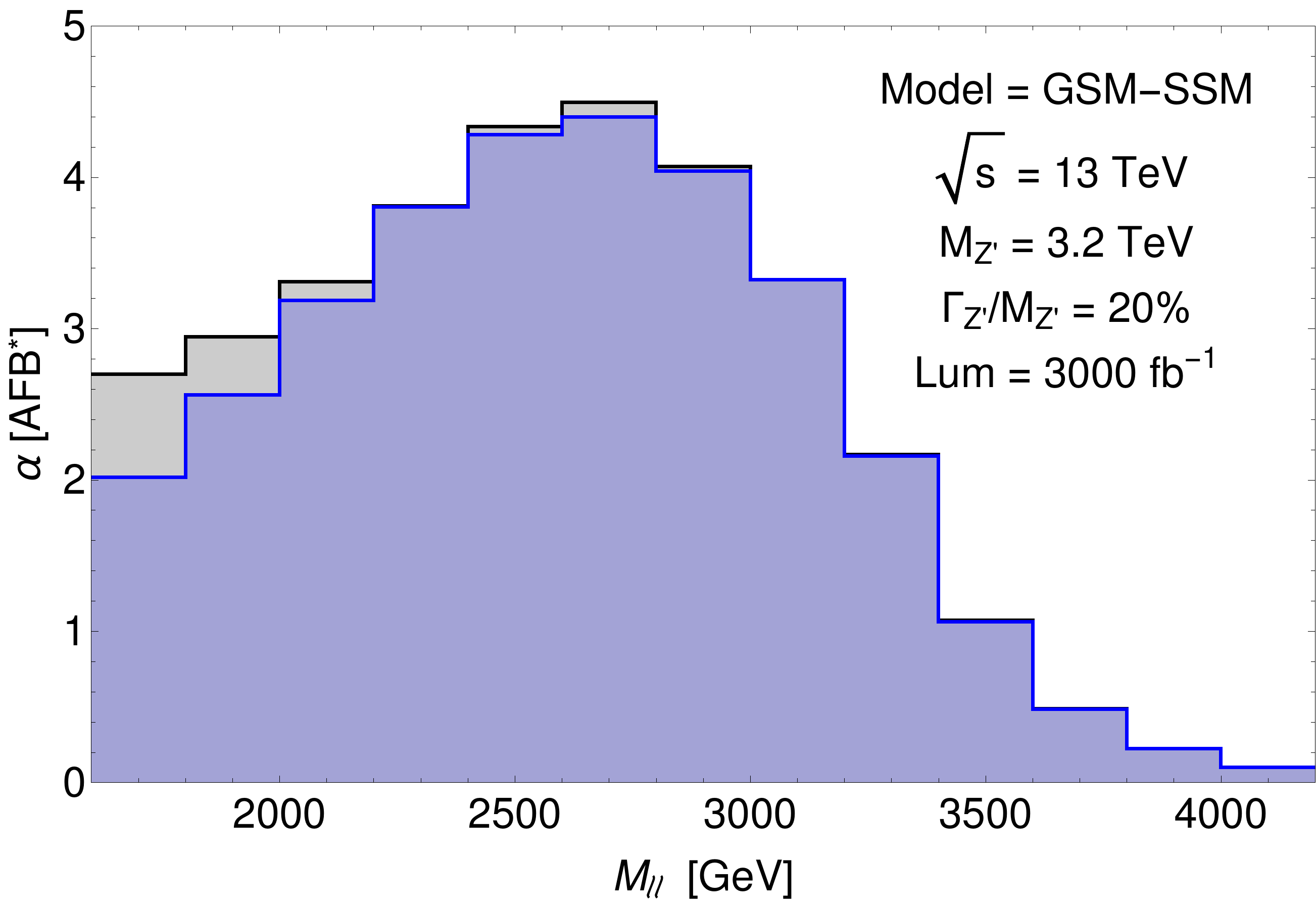}{(d)}
\caption{(a)Differential cross section significance within the GSM-SSM  model with a $Z^\prime$-boson of mass $M_{Z'}=$ 3 TeV and $\Gamma / M_{Z^\prime} = 20\%$. The black line represents the case where only the default DY process is accounted for as a SM background. 
The red line represents the case where the combined (DY + PI) process is taken into account as SM background. The integrated luminosity is $L=300~fb^{-1}$. 
(b) Same as (a) for a $Z^\prime$ with $M_{Z'}=$ 3.2 TeV and an integrated luminosity $L=3~ab^{-1}$.
(c) Significance of the reconstructed AFB distribution with and without the PI contribution, for a $Z^\prime$ boson of mass $M_{Z'}=$ 3 TeV and an integrated luminosity $L=300~fb^{-1}$. 
(d) Same as (c) for a $Z^\prime$ with $M_{Z'}=$ 3.2 TeV and an integrated luminosity $L=3~ab^{-1}$. Standard acceptance cuts are applied ($|\eta_l| < 2.5$ and $p_T^l > 20~$GeV) as well as the declared efficiencies of the electron and muon channels \cite{Khachatryan:2014fba}. NNLO QCD corrections are accounted for in the DY term \cite{Hamberg:1990np}. The overall significance is the combination of the significances in the two lepton channels. The binning has been chosen to represent an average of the two channel resolutions.} 
\protect{\label{fig:SSM_sig}}
\end{figure}

\noindent
As for the previous narrow $Z^\prime$-boson case, we calculate the significance of this type of BSM signal for the dilepton spectrum and the reconstructed AFB. 
The results are shown in Fig. \ref{fig:SSM_sig}. 
As before the black and the colored lines represent the projected result of a traditional analysis and the result obtained when one correctly includes the PI contribution in the SM background expectation and its PDF uncertainty in the total error. 
From the two upper plots in \ref{fig:SSM_sig}, we evince that the increase in significance expected at higher luminosities for a given mass $M_{Z'}$ is hampered by the presence of QED effects. 
The situation is much cleaner for the reconstructed AFB, which is very mildly affected by the uncertainties on the PI contribution.

\noindent
The broad resonance case shares similarities with the non-resonant case. 
In this instance, the theoretical interpretation of any excess of events in the dilepton spectrum would suffer the presence of large uncertainties in the SM background estimate. 
The accurate knowledge of the latter is in fact limited by the QED PDF uncertainties, especially when considering the usual dilepton spectrum as primary observable. 
The reconstructed AFB is clearly less affected by QED effects and, particularly in presence of either a broad resonance or a non resonant type of new physics, it could help validate a possible excess of events observed in the dilepton spectrum that would otherwise be very difficult to interpret owing to the large theoretical uncertainties.

\noindent
An advisable strategy would then be working with both observables. 
The spectrum should be used first to detect any excess, the AFB should intervene in the post discovery process of interpreting the obtained experimental results.

\section{Conclusions}\label{sec:summary}

In the search for dilepton signals of new physics that may appear at large invariant masses in LHC data samples one ought to rely upon a high level of control of the SM background, as possible manifestations of BSM signals may occur in observables in which the shape and size of signal and background are similar. 
This could be the case for both the cross section (when the signal of a, e.g., $Z^\prime$ state is very broad) and AFB, the latter having been hailed as a discovery variable (other than a diagnostic one) to be used alongside the former, precisely in such cases (of broad resonances).

\noindent
Hence, in this paper, we estimated the relative impact of the two dominant irreducible backgrounds from the SM onto dilepton searches at high invariant masses: (i) DY production via $s$-channel exchange of a $\gamma$ and $Z$ boson (both being off-shell); (ii) photo-production via PI in  $t,u$-channel scattering. 
In order to do so, we resorted to the only three available sets of PDFs including also photons in the evolution equations of partons inside the proton as well as deeply interacting states. 
The three choices were the most recent NNPDF set, the MRST2004QED distribution and the CT14QED set.

\noindent     
We found that, from the viewpoint of assessing theory systematics, while the DY channel is generally under control as far as the prediction of the central value and the associated error are concerned, this is no longer the case for the PI process, irrespectively of which PDF set is used. 
On the one hand, the NNPDF prescription for computing the PDF errors yields a large uncertainty, which can be of order ${\cal O}(100\%)$. On the other hand,  the MRST2004QED set offers no error analysis, 
while    the error extraction procedure for   the CT14QED  set  is not  yet systematic. 
Indeed, large differences occur between the three packages in the central value predictions as well. 
All such features are well visible in the cross section distribution mapped against the dilepton invariant mass.  

\noindent
One way to overcome the issue is trying to suppress the photon induced SM background to BSM searches, so that its theoretical uncertainty becomes implicitly harmless. 
We have analysed different sets of kinematical cuts that might be imposed experimentally. 
We have shown that the best option would be stiffening the selection of the dilepton final state by reducing the $|\eta_l|$ range of each lepton. 
In this manner, it is possible to reduce somewhat the contribution of the PI process relatively to the DY one, owing to the fact that the $t,u$-channel QED dynamics requires the final state leptons to be in the forward/backward direction more often than in the case of the DY channel. 
However, it remains unclear whether such cuts justify the loss of BSM signal, which also occurs.

\noindent
Another way of keeping under control the QED theoretical uncertainty is resorting to the reconstructed forward-backward charge asymmetry. 
The theoretical errors coming from the QED PDFs are in fact largely reduced in the case of AFB (again, plotted versus $M_{ll}$), thus reinforcing the use of this observable as a well motivated discovery probe. 
The reason is twofold. 
Firstly, being a ratio of cross sections, independently of the di-lepton production mode, systematic errors due to either PDF sets cancel out to a large extent. 
Secondly, the more troublesome of the two contributions (i.e. the PI one) does not contribute (at lowest order) to AFB, only to the overall normalization. 
We have illustrated this phenomenology for the case of both a narrow and wide $Z^\prime$-boson.

\noindent
Therefore, in absence of theoretical improvements in the modeling of the photon contribution to the evolution and extraction of PDFs, it becomes crucial to tension the results from the search for $Z^\prime$ states (or indeed any other type of new physics giving rise to lepton-pairs asymmetrically with respect to the beam direction) obtained from the dilepton spectrum against those coming from the reconstructed AFB.

\noindent
Our results on the relative contribution of PI processes in dilepton searches for new physics signals at the LHC warrant further, detailed investigations of precision Standard Model physics in the photon-photon channel. 
This includes higher-order corrections to the evolution of the photon PDF and to photon-initiated hard production processes, as well as the consistent treatment of contributions from finite photon virtualities in the high-energy collision. 
In this respect, we plan to investigate approaches to the equivalent photon approximation in a forthcoming study. 
We expect relevant Standard Model effects in searches based on PI processes to involve both QED/electroweak corrections and QCD contributions from the coupling of photons to jets.

\vskip0.35cm\noindent
\underbar{\sl Note added}.~The salient features of this work were presented by JF in the 
CMS public meeting~\cite{Fiaschi:2016tk} and 
by EA at MASS2016~\cite{Accomando:2016tk}. Further, while finalizing the present paper we became aware of Ref. \cite{Bourilkov:2016qum}, with which we agree in most respects.

\section*{Acknowledgements}
EA, JF and SM are financed in part through the NExT Institute. FH thanks the University of Hamburg and DESY for hospitality while part of this work was being done, and acknowledges partial support from the Collaborative Research Centre DFG SFB 676 ``Particles, Strings and the Early Universe".  We are grateful 
to   A.~Belyaev, J.~Huston,    A.~Pukhov and C.~Schmidt 
for  fruitful discussions.    We  thank 
M.~Dyndal and L.~Schoeffel for a precious consultation on the EPA contribution.

\bibliographystyle{apsrev4-1}
\bibliography{bib}

\end{document}